\documentclass[eqsecnum,superscriptaddress,nofootinbib,twocolumn]{revtex4-2}

\usepackage{graphicx} % for graphics
\usepackage{amsmath} % AMS math
\usepackage{amssymb} % AMS symbols
\usepackage{mathrsfs} % mathsfs font \mathscr{...}
\usepackage{bbm} % for fonts: \mathbbm{...} \mathbbmss{...} \mathbbmtt{...}; e.g. \mathbbm{1}
\usepackage{MnSymbol} % for using \largestar \medstar \smallstar
\usepackage{enumerate} % for using various item labels

\usepackage{footnotebackref} % enhanced hyperref (footnote back to the main text)

\newcommand{\abs}[1]{{\left|{#1}\right|}} % abs (variant delimiters)
\newcommand{\inner}[2]{{\langle {#1}\vert {#2} \rangle}} % inner product
\newcommand{\ket}[1]{\vert{#1}\rangle} % ket
\newcommand{\bra}[1]{\langle{#1}\vert} % bra
\newcommand{\vx}{\mathbf{x}} % vector of x
\newcommand{\vk}{\mathbf{k}} % vector of k
\newcommand{\kx}{\mathbf{k}\cdot\mathbf{x}} % k dot x
\newcommand{\vv}{\mathbf{v}} % vector of v

\newcommand{\secref}[1]{Sec.~\ref{#1}}
\newcommand{\appref}[1]{Appendix~\ref{#1}}
\newcommand{\figref}[1]{Fig.~\ref{#1}}
\newcommand{\footref}[1]{\textsuperscript{\ref{#1}}} % in case not defined in revtex4-2

\begin{document}
\count\footins = 1000 % prevents long footnotes from exceeding page bounds in revtex4-2

\title
{Two-path interference of single-particle pulses measured by the Unruh-DeWitt-type quantum detector}

%\thanks{Some comment here\dots}
\author{Bo-Hung Chen}
\email{kenny81778189@gmail.com}
\affiliation{Department of Physics, National Taiwan University, Taipei 10617, Taiwan}
\affiliation{Center for Theoretical Physics, National Taiwan University, Taipei 10617, Taiwan}

\author{Tsung-Wei Chen}
\email{twchen@mail.nsysu.edu.tw}
\affiliation{Department of Physics, National Sun Yat-sen University, Kaohsiung 80424, Taiwan}

\author{Dah-Wei Chiou}
\email{dwchiou@gmail.com}
\affiliation{Department of Physics, National Sun Yat-sen University, Kaohsiung 80424, Taiwan}
\affiliation{Center for Condensed Matter Sciences, National Taiwan University, Taipei 10617, Taiwan}

%\date{\today}

\begin{abstract}
We study the two-path interference of single-particle pulses measured by the Unruh-DeWitt-type quantum detector, which itself is a quantum state as well as the incoming pulse, and of which the interaction with the pulse is described by unitary quantum evolution instead of a nonunitary collapsing process. Provided that the quantum detector remains coherent in time long enough, the detection probability still manifests the two-path interference pattern even if the length difference between the two paths considerably exceeds the coherence length of the single-particle pulse, contrary to the result measured by an ordinary classical detector. Furthermore, it is formally shown that an ensemble of identical Unruh-DeWitt-type quantum detectors collectively behaves as an ordinary classical detector, if coherence in time of each individual quantum detector becomes sufficiently short. Our study provides a concrete yet manageable theoretical model to investigate the two-path interference measured by a quantum detector and facilitates a quantitative analysis of the difference between classical and quantum detectors. The analysis affirms the main idea of decoherence theory: quantum behavior is lost as a result of quantum decoherence.
\end{abstract}

%\pacs{???,???,???}

\maketitle

\section{Introduction}
Wave-particle duality is a central concept of quantum mechanics, which holds that every quantum object possesses properties of both waves and particles, appearing sometimes like a wave, sometimes like a particle, in different observational settings. Whether a quantum object is in the state of being a wave or a particle cannot be presupposed until it is measured.

\begin{figure}

\centering
   \includegraphics[width=0.4\textwidth]{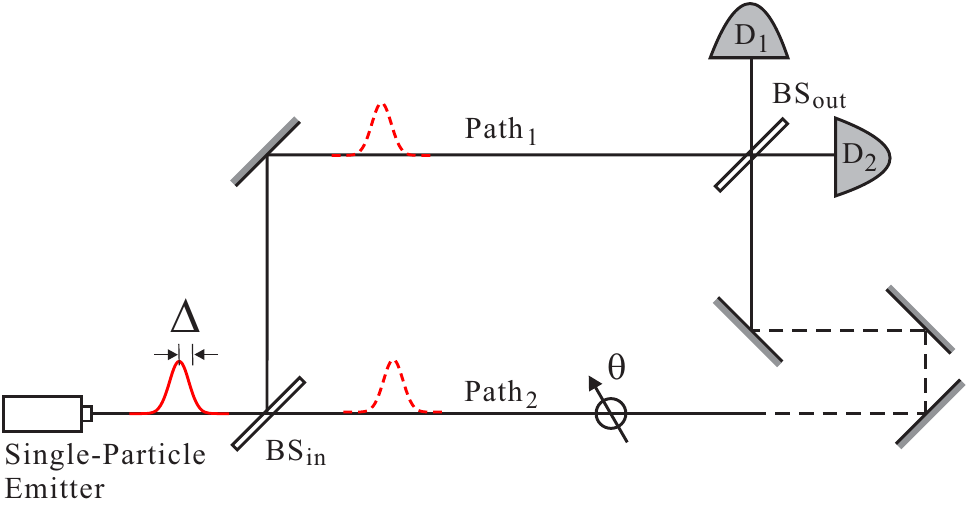}

\caption{Schematic plot of the Mach-Zehnder interferometer. Identical single-particle pulses with the spatial width $\Delta$ are fired from the single-particle emitter. The two paths of interference are labelled as $\mathrm{Path_1}$ and $\mathrm{Path_2}$, and their lengths are denoted as $L_1$ and $L_2$, respectively. A detour (dashed route) can be added to one of the two paths to adjust the length difference $\Delta L:=L_2-L_1$ between the two paths. A phase-shit plate is inserted to one of the two paths to provide an adjustable phase shit $\theta$. The two detectors detecting the single-particle pulse are denoted as $\mathrm{D_1}$ and $\mathrm{D_2}$; they can be either classical detectors or quantum detectors. In the case that $\mathrm{D_1}$ or $\mathrm{D_2}$ is a quantum detector, an additional classical detector (not shown here) has to be employed to obtain the final readout of the state of the quantum detector.}\label{fig:interferometer}
\end{figure}

To understand the wave-particle duality, consider a Mach-Zehnder interferometer as sketched in \figref{fig:interferometer}. A single-photon pulse is fired from the single-photon source and split by the first beam splitter $\mathrm{BS_{in}}$ into two paths. The two paths are recombined by a second beam splitter $\mathrm{BS_{out}}$ before the photon strikes either of the two detectors. The detection probabilities at the two detectors $\mathrm{D_1}$ and $\mathrm{D_2}$, which are measured as accumulated counts of signals of individual single-particle pulses, exhibit the wave nature of interference between the two paths $\mathrm{Path_1}$ and $\mathrm{Path_2}$ in the sense that they appear as modulated in response to an adjustable phase shift $\theta$.\footnote{The adjustable phase shift $\theta$ can be achieved, for example, by inserting a phase-shift plate into one of the two paths and tilting it with a piezoelectric actuator.}

On the other hand, if one manages to identify which path each individual photon travels through, the photons will exhibit the particle nature and the detection probabilities will be independent of $\theta$, showing no interference between the two paths.
The wave-particle duality is even more astonishing in Wheeler's delayed-choice experiment \cite{wheeler1979quantum}, where the choice of whether or not to measure ``which-path'' information of the photon can be made \emph{after} the photon has entered $\mathrm{BS_{in}}$, implying that a choice made in a later moment can \emph{retroactively} collapse a quantum state in the \emph{past}. This has been confirmed in various actual experiments \cite{alley1986results, hellmuth1987delayed, baldzuhn1989wave, lawson1996delayed, kim2000delayed, kawai1998realization, jacques2007experimental}.

However, if the difference between the length $L_1$ of $\mathrm{Path}_1$ and the length $L_2$ of $\mathrm{Path_2}$ becomes considerably larger than the spatial width $\Delta$ of the single-particle pulse, the pattern of modulation in response to $\theta$ becomes invisible, and apparently the interferometer ceases to exhibit the wave nature of photons regardless of whether the which-path information is identified or not.
In fact, the \emph{coherence length} $l_c$ of the source is defined as $\abs{\Delta L}:=\abs{L_2-L_1}$ for which the visibility of modulation is decreased by a certain factor compared to the case of $\Delta L=0$.\footnote{See Footnote \footref{foot:fringe visibility} and the context thereof for more details.}

In this paper, we argue that the reason why the interference disappears for large $\Delta L$ is essentially because the single-particle detector is or effectively behaves like an ordinary classical system, which has no or little coherence in time.
If the classical detector is replaced by a quantum detector that can remain coherent in time long enough, the detection probability by the quantum detector will still manifest the two-path interference even for large $\Delta L$.

An ordinary classical detector is different from a quantum detector essentially in two aspects. First, whereas the interaction between the to-be-measured quantum state and the classical detector is described by a nonunitary collapsing process, the interaction between the to-be-measured state and the quantum detector is described by a unitary quantum evolution. Second, whereas the classical detector yields the readout of the to-be-measured quantum state immediately upon its interaction with the quantum state, the quantum detector does not yield the readout until an additional classical detector is employed to measure the status of the quantum detector, from which the final readout of the to-be-measured state is inferred. The quantum collapse of the whole system of the to-be-measured state and the quantum detector takes place at the moment when the final measurement by the additional classical detector is performed.\footnote{\label{foot:clarification}Opinions differ as to what kinds of detection processes should be deemed ``classical'' or ``quantum'', as a measurement in general may be mediated via ancillary (quantum) degrees of freedom, while the final outcome is always read out classically. In this paper, a ``classical detector'' of single-particle pulses refers to an \emph{ordinary} detector that registers a signal whenever the incoming pulse strikes it, regardless of the arrival time. A photographic film and a charge-coupled device (CCD) are two typical examples. Even though the underlying mechanism of the interaction between the incoming pulse and the classical detector is essentially quantum (of course), the exact form of the mechanism is unimportant and needs not to be modeled, as the detection is described by a nonunitary collapsing process. On the other hand, a ``quantum detector'' of single-particle pulses refers to a quantum system which interacts unitarily with the incoming pulse and of which the quantum state indicates whether a particle is detected. The Unruh-WeWitt detector provides such an example. This paper does not intend to define the classical-quantum dichotomy; nor does it consider general measurement invoking arbitrary ancillary degrees of freedom. The adjectives ``classical'' and ``quantum'' are used for the specific meanings as stated above.}

For a quantum detector that remains coherent in time long enough, the arrival time at which the incoming pulse strikes the detector is unknowable (within the coherence time of the detector). According to the ``sum-over-histories'' principle in quantum mechanics, one should sum amplitudes \emph{interferentially} over all possible values of the arrival time (within the coherence time of the detector) to obtain the probability that the detector is said to register a signal upon the final measurement by the additional classical detector. Therefore, it is anticipated that the detection probability still manifests the two-path interference pattern, even if the length difference $\Delta L$ considerably exceeds the coherence length $l_c$.
By contrast, as a classical detector has no or little coherence in time, its detection probability is obtained by summing probabilities \emph{additively} over all possible values of the arrival time (i.e., \emph{no} quantum interference \emph{over time}), even if the arrival time is not explicitly measured. Consequently, the two-path interference pattern will diminish when $\Delta L$ exceeds $l_c$.\footnote{\label{foot:distinction}The distinction between the quantum and classical detection probabilities is \emph{not} the ability of the detector to resolve the arrival time. For a CCD, the arrival time can be measured with high resolution if it is coupled to a high-precision clock. However, whether the arrival time is measured or not, the detection probability of a CCD remain the renowned classical pattern, i.e., as given in \eqref{classical P result}. For a photographic film, there is even no obvious way to measure the arrival time with high resolution, yet it gives the same classical probability pattern. Both classical and quantum detectors can be ignorant of the arrival time. In a sense, there are two distinct kinds of ignorance of the arrival time: ``quantum ignorance'' and ``classical ignorance''. The former gives quantum interference over time, while the latter does not. As this paper will demonstrate, the distinction is whether the detector can remain \emph{coherent} enough in time, not whether it remains \emph{ignorant} enough of time.}

Although the arguments given above are sensible, the anticipated results have not been explicitly demonstrated in a well-posed model, mainly because it is rather difficult to devise a concrete yet manageable model of a quantum detector for measuring the two-path interference.
In this paper, adopting the idea of the Unruh-DeWitt detector \cite{unruh1976notes, dewitt1979quantum, Barbado:2020snx, Foo:2020xqn, Martin-Martinez:2012ysv} primarily used in the literature of quantum field theory in curved spacetime (for reviews, see \cite{birrell1984quantum,wald1994quantum,padmanabhan2005gravity} and especially \cite{crispino2008unruh}), we formulate the quantum detector and its interaction with a single-particle state in the style of an Unruh-DeWitt detector.\footnote{To be as generic as possible, we consider single-particle pulses not only of photons but also of other massless or massive matter fields.} This provides a manageable theoretical model to investigate the two-path interference measured by a quantum detector and facilitates a quantitative analysis.

The quantitative analysis explicitly shows that the detection probability measured by an Unruh-DeWitt-type quantum detector manifests the two-path interference pattern even if $\Delta L$ considerably exceeds $l_c$, provided that the quantum detector remains coherent in time long enough.
Furthermore, by formally deforming the switching function, which accounts for the switch-on period of the Unruh-DeWitt detector, we are able to model a quantum detector with a finite period of coherence in time and carry out a quatitative investigation of the difference between classical and quantum detectors. The investigation formally shows that a large ensemble of identical Unruh-DeWitt-type quantum detectors collectively behaves as a single ordinary classical detector if coherence in time of each individual quantum detector becomes sufficiently short.
Although our model only formally takes into account quantum decoherence without formulating its underlying mechanism, the result of our model nevertheless affirms the main idea of \emph{decoherence theory} (see \cite{schlosshauer2005decoherence} for a review): \emph{quantum behavior is lost as a result of quantum decoherence.}
It might also shed new light on the measurement problem in quantum mechanics. Particularly, it seems to support the interpretation of quantum collapse in \emph{objective-collapse theories} (e.g.\ see \cite{ghirardi1985model, ghirardi1986unified, ghirardi1990markov}, \cite{penrose1996gravity, penrose1998quantum, penrose2014gravitization}, and \cite{jabs2016conjecture} for different theories), which holds that a quantum state in superposition is collapsed (localized) spontaneously when a certain objective physical threshold is reached.

\section{Single-particle pulse states}
Before investigating the two-path interference, we first give a mathematical description of the single-particle pulse and its propagation along the two paths. We assume that the detectors are agnostic of polarization or spin, and therefore, for simplicity, the pulse is modeled in terms of a real scalar (Klein-Gordon) field:
\begin{eqnarray}\label{phi x t}
&& \phi(\vx,t) = e^{iH_\phi t} \phi(\vx) e^{-iH_\phi t} \\
&=& \int \frac{d^nk}{(2\pi)^n} \frac{1}{\sqrt{2\omega_\vk}}
\left(a_\vk e^{i(\kx-\omega_\vk t)}+a_\vk^\dag e^{-i(\kx-\omega_\vk t)}\right),\nonumber
\end{eqnarray}
where $\phi(\vx):=\phi(\vx,t=0)$, $H_\phi$ is the Hamiltonian of the field $\phi$, and
\begin{equation}
\omega_\vk = \sqrt{\vk^2+m^2}
\end{equation}
is the energy (frequency) corresponding to $\vk$. To make the formulation as generic as possible, we begin with it in $n+1$ dimensions and consider both the cases of $m=0$ (e.g.\ photons) and $m\neq0$ (i.e.\ massive matter fields).

The initial state of a single-particle pulse emitted by the single-particle emitter can be described as a superposition of single-particle states given by
\begin{equation}\label{Psi0}
\ket{\Psi_0} = \int d^nx f(\vx) \phi(\vx)^\dag \ket{0}
= \int d^nx f(\vx) \phi(\vx) \ket{0},
\end{equation}
where $f(\vx)$ is a wave packet describing the profile of the pulse.
Particularly, $f(\vx)$ can be modeled as
\begin{equation}\label{wave packet}
f(\vx) = e^{i\vk_0\cdot\vx}\prod_{i=1}^{n}\frac{e^{-x_i^2/(2\Delta_i)^2}}{(2\pi\Delta_i^2)^{1/4}},
\end{equation}
which is a plane wave enveloped by a Gaussian wave packet function that is centered at $\langle x_i\rangle:=\int_{-\infty}^\infty d^nx\, x_i f(\vx)^*f(\vx)=0$ with a spatial width $\langle\Delta x_i\rangle^2 \equiv \langle x_i^2\rangle-\langle x_i\rangle^2 :=\int_{-\infty}^\infty d^nx\, x_i^2 f(\vx)^*f(\vx) - \langle x_i\rangle^2 = \Delta_i^2$ in each spatial direction.
Meanwhile,
\begin{equation}
\phi(\vx)\ket{0} = \int\frac{d^nk}{(2\pi)^n}\frac{1}{2\omega_\vk} e^{-i\kx}\ket{\vk},
\end{equation}
where we have defined
\begin{equation}
\ket{\vk}:=\sqrt{2\omega_\vk}\,a_\vk^\dag\ket{0},
\end{equation}
which satisfies
\begin{equation}
\inner{\vk}{\vk'} = 2\omega_\vk(2\pi)^n\delta^{(n)}(\vk-\vk')
\end{equation}
and is interpreted as a single-particle momentum state.
The state $\phi(\vx)\ket{0}$ is interpreted as having a single particle at position $\vx$, as it can be shown $\bra{0}\phi(\vx)\ket{\vk}\equiv \bra{0}\phi(\vx)\sqrt{2\omega_\vk}\,a_\vk^\dag\ket{0} =e^{i\kx}$.\footnote{For more detailed interpretations of $\ket{\vk}$ and $\phi(\vx)\ket{0}$, see Chapter 2 of \cite{peskin2018introduction}.}
Therefore, the state $\ket{\Psi_0}$ given by \eqref{Psi0} can be understood as a single-particle state whose probability of being found at position $\vx$ is specified by the pulse amplitude $f(\vx)$.

The Fourier transform of \eqref{wave packet} is given by
\begin{eqnarray}\label{f tilde}
\tilde{f}(\vk) &:=& \left(\frac{1}{\sqrt{2\pi}}\right)^n \int_{-\infty}^\infty d^nx f(\vx)e^{-i\kx} \nonumber\\
&=& \left(\prod_{i=1}^n\frac{2\Delta_i^2}{\pi}\right)^{1/4} e^{-(k_i-k_{0i})^2\Delta_i^2},
\end{eqnarray}
and note that $\tilde{f}(\mathbf{k})=\tilde{f}(\mathbf{k})^*$.
The state $\ket{\Psi_0}$ can be recast in the $\vk$ space as
\begin{equation}\label{Psi0 in k}
\ket{\Psi_0} = \int\frac{d^nk}{(2\pi)^n}\frac{\left(\sqrt{2\pi}\right)^n}{2\omega_\vk} \tilde{f}(\vk) \ket{\vk}.
\end{equation}
The spectrum in the $\vk$ space is given by a Gaussian distribution centered at $\langle \vk\rangle:=\int_{-\infty}^\infty d^nk\, \vk \tilde{f}(\vk)^*\tilde{f}(\vk)=\vk_0$ with an uncertainty $\langle\Delta k_i\rangle^2 \equiv \langle k_i^2\rangle-\langle k_i\rangle^2 :=\int_{-\infty}^\infty d^nk\, k_i^2 \tilde{f}(\vk)^*\tilde{f}(\vk) - \langle k_i\rangle^2 = 1/(2\Delta_i)^2$ in each direction.\footnote{Note that the relativistic factor $1/\sqrt{2\omega_\vk}$ in \eqref{Psi0 in k} can be treated as nearly constant, provided the profile $\tilde{f}(\vk)$ is sharp enough. For more details of this factor, see Section 2.5 of \cite{peskin2018introduction}.}
Note that the wave packet given by \eqref{wave packet} saturates the uncertainly principle, i.e., $\langle\Delta x_i\rangle \langle\Delta k_i\rangle = 1/2$.

By the identities $e^{-iH_\phi t}a_\vk e^{iH_\phi t}=a_\vk e^{i\omega_\vk t}$, $e^{-iH_\phi t}a_\vk^\dag e^{iH_\phi t}=a_\vk^\dag e^{-i\omega_\vk t}$, and $e^{-iH_\phi t}\ket{0}=\ket{0}$, the initial state \eqref{Psi0} evolves into
\begin{subequations}\label{Psi t}
\begin{eqnarray}
\label{Psi t a}
&&\ket{\Psi(t)} = e^{-iH_\phi t}\ket{\Psi_0} \nonumber\\
&=& \int d^nx f(\vx) \int\frac{d^nk}{(2\pi)^n} \frac{e^{-i\kx-i\omega_\vk t}}{\sqrt{2\omega_\vk}} a_\vk^\dag\ket{0} \\
\label{Psi t b}
&\equiv& \int\frac{d^nk}{(2\pi)^{n/2}}\frac{\tilde{f}(\vk)}{\sqrt{2\omega_\vk}}\, a_\vk^\dag e^{-i\omega_\vk t}\ket{0}
\end{eqnarray}
\end{subequations}
at time $t$.
Because the function $\tilde{f}(\vk)$ is sharply centered at $\vk=\vk_0$, it is a good approximation if we only take into account those $\mathbf{k}$ close to $\vk=\vk_0$ in \eqref{Psi t}. Correspondingly, we take the Taylor expansion of $\omega_\vk$ around $\vk=\vk_0$:
\begin{eqnarray}\label{Taylor expansion}
\omega_\vk &=& \omega_\vk\big|_{\vk=\vk_0} + \boldsymbol{\nabla}_\vk\omega_\vk\big|_{\vk=\vk_0}\!\! \cdot (\vk-\vk_0) + O((\vk-\vk_0)^2) \nonumber\\
&\equiv& \omega_0 + \vv_0\cdot(\vk-\vk_0) + O((\vk-\vk_0)^2),
\end{eqnarray}
where we have defined the energy at $\vk=\vk_0$ as
\begin{equation}\label{omega0}
\omega_0 := \omega_\vk\big|_{\vk=\vk_0},
\end{equation}
and the group velocity at $\vk=\vk_0$ as
\begin{equation}
\vv_0 := \boldsymbol{\nabla}_\vk\omega_\vk\big|_{\vk=\vk_0}.
\end{equation}
Substituting \eqref{Taylor expansion} into \eqref{Psi t a}, we have (see \appref{app:Psi t 2} for the detailed derivation)
\begin{equation}\label{Psi t 2}
\ket{\Psi(t)} \approx  e^{-i\delta\omega_0t}\int d^nx f(\vx-\vv_0t) \phi(\vx)\ket{0},
\end{equation}
where we have defined
\begin{equation}
\delta\omega_0 := \omega_0-\vv_0\cdot\vk_0.
\end{equation}
Compared with \eqref{Psi0}, the wave function of \eqref{Psi t} is identical to $\ket{\Psi_0}$, except that it is translated by $\vv_0t$ (and multiplied by a phase factor $e^{-i\delta\omega_0t}$). This is expected, since the approximation \eqref{Taylor expansion} essentially neglects the effect of dispersion and correspondingly the wave packet's profile remains undeformed while propagating with the group velocity $\vv_0$.\footnote{Note that, in the case of $m=0$, the approximation of no dispersion becomes exact, since the part $O((\vk-\vk_0)^2)$ in \eqref{Taylor expansion} vanished identically.}

Now, consider that the initial state propagates along $\mathrm{Path}_1$ or $\mathrm{Path}_2$ and is finally interacted with $\mathrm{D}_1$ or $\mathrm{D}_2$. Since the lateral degrees of freedom perpendicular to the propagating direction are inessential as regards the two-path interference, we can treat the propagation along each path as a $1+1$ dimensional problem (i.e., $n=1$), and denote the spatial width $\Delta_i$ of the single-particle pulse in the propagating direction simply as $\Delta$.\footnote{The approximation by disregarding lateral degrees is legitimate if the \emph{lateral coherence} of the source is sufficiently good or the lateral dimension of the propagating pulse remains sufficiently small. See Section 2.9 of \cite{menzel2013photonics} for more details.}
First, we study the case along $\mathrm{Path}_1$ and interacted with $\mathrm{D}_1$.
Let $\mathrm{D}_1$ located at $x=L_1$ (with respect to the emitter at $x=0$ along $\mathrm{Path}_1$) and $y:=x-L_1$  be the relative location with reference to $\mathrm{D}_1$. In terms of $y$, it follows from \eqref{Psi t 2} that
\begin{eqnarray}\label{path1 state}
&&\ket{\Psi(t)} \nonumber\\
&\approx&
e^{-i\delta\omega_0t}
\int dy f(L_1+y-v_0t) \!\int \frac{dk}{2\pi}\frac{e^{-ik(L_1+y)}}{\sqrt{2\omega_k}}a_k^\dag\ket{0}\nonumber\\
&\approx& e^{-i\delta\omega_0t} e^{-ik_0L_1}
\int dy f(L_1+y-v_0t) \phi(y)\ket{0},
\end{eqnarray}
where we have made another approximation $e^{-ikL_1}\approx e^{-ik_0L_1}$ to move the factor $e^{-ikL_1}$ out of the integral on the grounds that the factor $\int dy f(L_1+y-v_0t) e^{-iky}=\tilde{f}(k)e^{-ik(L_1-v_0t)}$ is sharply centered around $k=k_0$ by the profile of $\tilde{f}(k)$.

Similarly, in terms of the relative location $y$ with reference to $\mathrm{D}_1$, the wave function along $\mathrm{Path}_2$ is given by
\begin{eqnarray}\label{path2 state}
\ket{\Psi(t)} &\approx&
e^{-i\delta\omega_0t} e^{-i(k_0L_2-\theta)}
\int dy f(L_2+y-v_0t)\phi(y)\ket{0},\nonumber\\
\end{eqnarray}
where an extra phase $e^{i\theta}$ is added to account for the adjustable phase shift.
The wave functions along the two paths are recombined by $\mathrm{BS}_\mathrm{out}$. As a result, in terms of the relative coordinate $y$ with reference to $\mathrm{D}_1$, the recombined wave function entering $\mathrm{D}_1$ is given by
\begin{equation}\label{Psi D1}
\ket{\Psi^{\mathrm{D}_1}(t)}
\approx e^{-i\delta\omega_0t} \int dy F_{\mathrm{D}_1}(y,t)
\phi(y)\ket{0},
\end{equation}
with
\begin{eqnarray}\label{F D1'}
F_{\mathrm{D}_1}(y,t) &=& \frac{e^{-ik_0L_1}}{2}f(L_1+y-v_0t) \\
&& \mbox{} + e^{ik_0\Delta L}\frac{e^{-i(k_0L_2-\theta)}}{2}f(L_2+y-v_0t), \nonumber
\end{eqnarray}
where the factor $1/2$ accounts for the assumption that $\mathrm{BS}_\mathrm{in}$ splits the incoming pulse equally to $\mathrm{Path}_1$  and $\mathrm{Path}_2$ (thus giving rise to a factor of $1/\sqrt{2}$) and then $\mathrm{BS}_\mathrm{out}$ splits the pulse coming from $\mathrm{Path}_1$ or $\mathrm{Path}_2$ equally to $\mathrm{D}_1$ and $\mathrm{D}_2$ (thus giving another factor of $1/\sqrt{2}$), and furthermore an extra phase $e^{ik_0\Delta L}$ is added to the wave function traveling along $\mathrm{Path}_2$. Because the two wave functions travel along \emph{different} lengths but the recombined result is cast in terms of the \emph{same} coordinate $y$, we have to add the extra phase $e^{ik_0\Delta L}$ to compensate for the artifact of a relative phase difference between the two paths resulting from combining the two paths of different lengths to the same coordinate $y$. Since the constant overall phase $e^{-ik_0L_1}$ in \eqref{F D1'} is of no physical significance, it is equivalent to express $F_{\mathrm{D}_1}(y,t)$ as
\begin{equation}\label{F D1}
F_{\mathrm{D}_1}(y,t) = \frac{1}{2}f(L_1+y-v_0t) + \frac{e^{i\theta}}{2}f(L_2+y-v_0t).
\end{equation}
Similarly, in terms of the relative coordinate $y$ with reference to $\mathrm{D}_2$ , the wave function entering $\mathrm{D}_2$ is given by
\begin{equation}\label{Psi D2}
\ket{\Psi^{\mathrm{D}_2}(t)}
\approx e^{-i\delta\omega_0t} \int dy F_{\mathrm{D}_2}(y,t)
\phi(y)\ket{0},
\end{equation}
with
\begin{equation}\label{F D2}
F_{\mathrm{D}_2}(y,t) = \frac{1}{2}f(L_1+y-v_0t) - \frac{e^{i\theta}}{2}f(L_2+y-v_0t).
\end{equation}
In comparison with $F_{\mathrm{D}_1}$, we have inserted a relative negative sign between the two paths for $F_{\mathrm{D}_2}$, accounting for the fact that reflection off the different sides of a beam splitter gives rise to an extra phase shift of $0$ or $\pi$, respectively.\footnote{If the source is not photons but other matter fields, different apparatuses (in replacement of the mirrors, beam splitters, and phase shifter) are used to realize the two-path interference. In this case, the extra phase shift may not be $0$ or $\pi$; nevertheless, our central conclusion that the detection probability is modulated in response to $\theta$ remains unchanged.}

\begin{figure*}

\centering

  \begin{minipage}[b]{0.48\textwidth}
    \includegraphics[width=\textwidth]{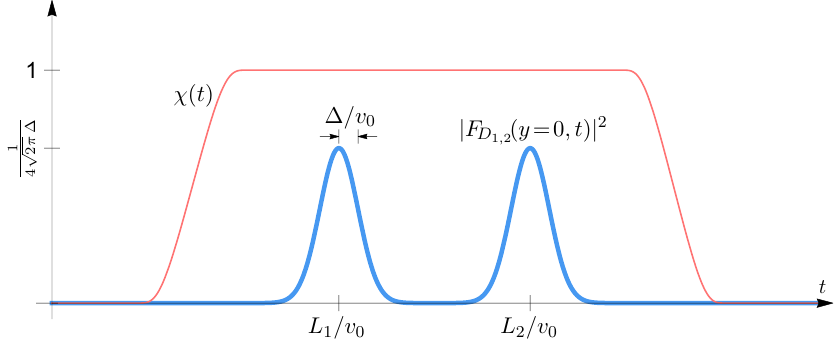}
    %\caption{part one.}
  \end{minipage}
  \hspace{0.2cm} %\hfill
  \begin{minipage}[b]{0.48\textwidth}
    \includegraphics[width=\textwidth]{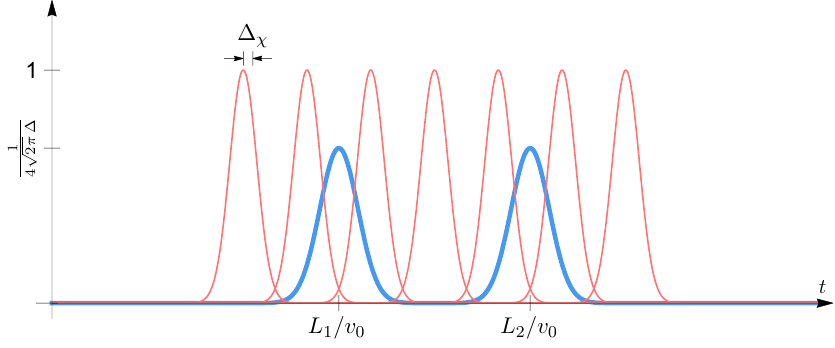}
    %\caption{part two}
  \end{minipage}

\caption{\textit{Left}: Schematic representation of the squared amplitude $\abs{F_{\mathrm{D}_{1,2}}(y,t)}^2$ measured at $y=0$ as a function of $t$ (thicker curve) and the switching function $\chi(t)$ (thiner curve). $F_{\mathrm{D}_{1,2}}(y,t)$ are given by \eqref{F D1} and \eqref{F D2}, and $\Delta$ is the spatial width of the single-particle pulse appearing in \eqref{wave packet} (for $n=1$), which is manifested as a temporal width $\Delta/v_0$ for the two wave packages in $\abs{F_{\mathrm{D}_{1,2}}(y=0,t)}^2$. Here, $\abs{F_{\mathrm{D}_{1,2}}(y=0,t)}^2$ is depicted for the case that $\abs{\Delta L}:=\abs{L_2-L_1}\gtrsim 6\Delta$, whereby the two wave packets are well separated and virtually have no overlap. The switching function $\chi(t)$ appearing in \eqref{H int} is depicted for the case that the switch-on period is long enough so that both wave packets are well covered.
\textit{Right}:
The switching function $\chi(t)$ is given by  the Gaussian form in \eqref{chi}. The parameter $\Delta_\chi$ for the switch-on period is depicted to be shorter than $\Delta L/v_0$. Many switching functions with the same $\Delta_\chi$ but different values of $t_\chi$ are shown together to illustrate the \emph{ensemble average} as given by \eqref{P tilde}.}\label{fig:functions}
\end{figure*}

The wave function $F_{\mathrm{D}_{1,2}}(y,t)$ given by \eqref{F D1} and \eqref{F D2} is simply the superposition of two identical propagating Gaussian wave packets multiplied by the phases $e^{0}$ and $\pm e^{i\theta}$, respectively, and separated by $\Delta L:=L_2-L_1$ in $y$.
\figref{fig:functions} illustrates its signature by depicting $\abs{F_{\mathrm{D}_{1,2}}(y,t)}^2$ as measured at $y=0$ as a function of $t$. If $\Delta L$ is much larger than the spatial width $\Delta$ of the wave packet (more precisely, $\Delta L\gtrsim 6\Delta$), the overlap of the two Gaussian packets and consequently the dependence on $\theta$ are virtually negligible as far as $\abs{F_{\mathrm{D}_{1,2}}(y,t)}^2$ is concerned.
This is essentially the reason why the interference in response to $\theta$ disappears for the case $\Delta L\gtrsim 6\Delta$, when measured by classical detectors.
By contrast, when measured by quantum detectors, the interference in response to $\theta$ can still be manifested, even if $\Delta L$ is considerably larger than $6\Delta$.
In the following, we will study the cases of classical and quantum detectors, respectively.

\section{Classical detectors}
A detector is said to be classical if it is not described as a quantum state but rather serves as a measuring device --- i.e., as long as the detector registers a signal, it induces quantum collapse upon the quantum state to be measured.
The interaction between the to-be-measured state and the classical detector is \emph{not} described by unitary quanftum evolution, but a nonunitary collapsing process.

Coherence in time of the classical detection is virtually negligible. That is, the events of registering a signal at two different moments $t$ and $t'$ are regarded as independent of each other. Therefore, the total probability of detecting a signal of a particle by the classical detectors $\mathrm{D}_1$ and $\mathrm{D}_2$ is simply the sum of probabilities of detecting a signal over all possible moments, i.e.,
\begin{equation}\label{classical P}
\mathcal{P}_{\mathrm{D}_{1,2}}^\mathrm{(cl)} \propto \int_{-\infty}^\infty \abs{F_{\mathrm{D}_{1,2}}(y=0,t)}^2 dt
\end{equation}
up to an overall numerical factor depending on the efficiency of detection.
There is \emph{no} quantum interference \emph{over time}.
Substituting \eqref{F D1} and \eqref{F D2} into \eqref{classical P} and performing the change of variables $x=L_1-v_0t$, we have
\begin{eqnarray}\label{classical P result}
  \mathcal{P}_{\mathrm{D}_{1,2}}^\mathrm{(cl)}
  &= &  \frac{1}{4} \int_{-\infty}^\infty dx\bigg( \abs{f(x)}^2
              \pm e^{-i\theta}f(x)f^*(x+\Delta L)\nonumber\\
  && \quad\mbox{} + \abs{f(x+\Delta L)}^2
                  \pm e^{i\theta}f^*(x)f(x+\Delta L)\bigg)  \nonumber\\
  &=& \frac{1}{4}\bigg(2\pm2\cos(k_0\Delta L+\theta) \nonumber\\
  && \qquad\qquad\quad \times
      \int_{-\infty}^\infty dx\frac{e^{-[(x+\Delta L)^2+x^2]/(2\Delta)^2}}{(2\pi\Delta^2)^{1/2}}\bigg) \nonumber\\
  &=& \frac{1}{2}\left(1\pm\cos(k_0\Delta L+\theta)\,
  e^{-\left(\frac{\Delta L}{2\sqrt{2}\Delta}\right)^2} \right),
\end{eqnarray}
where the explicit form of \eqref{wave packet} has been used and the overall numerical factor has been fixed by assuming perfect efficiency of detection.
In the case of $\Delta L\ll \Delta$, it reduces to
\begin{eqnarray}\label{cos sin modulation}
\mathcal{P}_{\mathrm{D}_{1,2}}^\mathrm{(cl)}
&=&
\frac{1}{2}\left(1\pm\cos(k_0\Delta L+\theta)\right) \nonumber\\
&=&
\left\{\begin{array}{l}
\cos^2\left((k_0\Delta L+\theta)/2\right), \\
\sin^2\left((k_0\Delta L+\theta)/2\right),
\end{array}
\right.
\end{eqnarray}
which is the renowned modulation of the two-path interference in response to $\theta$ plus the effective phase shift $k_0\Delta L$ due to the length difference.

On the other hand, if $\Delta L$ is comparable to $\Delta$, the interference pattern as modulated in response to $\theta$ will diminish due to the exponential factor of $e^{-(\Delta L/2\sqrt{2}\Delta)^2}$. In case $\Delta L$ is much larger than $\Delta$ (more precisely, $\Delta L\gtrsim 6\Delta$), the interference pattern completely dims out and we simply have $\mathcal{P}\approx50\%$ for each of $\mathrm{D}_1$ and $\mathrm{D}_2$.
The \emph{coherence length} $l_c$ of a light source is defined as $\abs{\Delta L}$ for which the visibility of modulation is decreased by a factor of $1/\sqrt{2}$, $1/2$, or $1/e$ (by different conventions) compared to the case of $\Delta L=0$.\footnote{\label{foot:fringe visibility}The coherence length $l_c$ of a pulse is an indicator of its \emph{longitudinal} coherence (in contrast to its \emph{lateral} coherence). The visibility of modulation (also referred to as \emph{fringe visibility}) is defined as $(\mathcal{P}_{\max}-\mathcal{P}_{\min})/(\mathcal{P}_{\max}+\mathcal{P}_{\min})$, where $\mathcal{P}_{\max}=\max_\theta \mathcal{P}(\theta)$ and $\mathcal{P}_{\min}=\min_\theta \mathcal{P}(\theta)$.} Obviously, up to a numerical factor of order $O(1)$, the coherence length $l_c$ is given by the spatial width $\Delta$ of the pulse, i.e.
\begin{equation}\label{lc}
l_c \sim \Delta \sim \frac{1}{2\Delta k} \sim \frac{1}{4\pi}\frac{\lambda_0^2}{\Delta\lambda},
\end{equation}
where $\lambda_0:=2\pi/k_0$ and $\Delta\lambda$ is the spectral uncertainty in wavelength.
The coherence length $l_c$ can also be understood as given by the inverse of the bandwidth $\Delta k$ (up to a numeral factor).
For more details of defining and measuring the coherence length, see Section 2.9 of \cite{menzel2013photonics}.\footnote{For a continuous wave source with the spectrum given by the same profile $\tilde{f}(k)$, the uncertainty $\Delta\omega\sim v_0\Delta k$ in frequency gives rise to phase drift in time. As a result, the continuous wave has the coherence time $\tau_c$ given by the uncertainty relation $\tau_c\Delta\omega\sim1/2$, which in turn gives the coherence length $l_c\sim\tau_c v_0\sim 1/(2\Delta k)$. The coherence length of a continuous wave is the same as that of a single-particle pulse as given in \eqref{lc}, although their interpretations are understood differently. This is expected, because the interference pattern of a continuous wave is supposed to be identical to the accumulated counts of individual single-particle experiments.}

\section{Unruh-DeWitt-type quantum detectors}\label{sec:quantum detectors}
Contrary to classical detectors, a detector is said to be quantum if the detector itself is described as a quantum state as well as the incoming pulse to be measured. The interaction between the to-be-measured quantum state and the quantum detector is governed by unitary quantum evolution. That is, the interaction between the to-be-measured and the detector does \emph{not} induce quantum collapse. Instead, the quantum collapse of the whole system of the to-be-measured and the detector takes place at the moment when an additional classical detector is employed to read out the state of the quantum detector. Before the final read-out, the quantum detector remains coherent in time. Therefore, as long as the coherence of the detector is not disturbed, the relative phase between the two wave packets can still be manifested via the quantum interaction, even if the two wave packets are well separated as depicted in \figref{fig:functions}.

The simplest way to model a single-particle quantum detector is to formulate it as an Unruh-DeWitt detector \cite{unruh1976notes, dewitt1979quantum, Barbado:2020snx, Foo:2020xqn, Martin-Martinez:2012ysv, birrell1984quantum,wald1994quantum,padmanabhan2005gravity,crispino2008unruh}, which is an idealized point-particle detector with two energy levels $\ket{E_0}$ and $\ket{E}$, coupled to a scalar field $\phi$ via a monopole interaction.
The energy gap between the two levels is denoted as $\Delta E:=E-E_0$.\footnote{We consider both the cases of $\Delta E>0$ and $\Delta E<0$ as commonly studied in the literature of the Unruh radiation (see e.g.\ \cite{crispino2008unruh}). In the end, as will be seen in \eqref{Fermi's golden rule}, only the case $\Delta E>0$ yields appreciable transition probability.}
If the detector moves along a trajectory $x^\mu(\tau)$ with $\tau$ being the detector's proper time, the monopole interaction is given by the interaction Hamiltonian (see\cite{unruh1976notes, dewitt1979quantum, Barbado:2020snx, Foo:2020xqn, Martin-Martinez:2012ysv, birrell1984quantum,wald1994quantum,padmanabhan2005gravity,crispino2008unruh})
\begin{equation}\label{H int}
H_\mathrm{int} = \kappa\, \chi(\tau)\mu(\tau)\phi(x^\mu(\tau)),
\end{equation}
where $\kappa$ is a small coupling constant, $\mu(\tau)$ is the operator of the detector's monopole moment, and $\chi(\tau)$ is a switching function. The switching function accounts for the switch-on and switch-off of the interaction; it is typically modeled as a smooth function of $\tau$ that increases from 0 to 1, then remains to be 1 for a certain period, and finally decreases from 1 to 0, as depicted on the left of \figref{fig:functions}.
In the laboratory frame, we put the Unruh-DeWitt detector at a fixed place, i.e.\ $x^\mu(\tau) = (y=0,t=\tau)$. The detector is said to register a signal of a single particle, if the detector happens to be excited ($\Delta E>0$) or deexcited ($\Delta E<0$) from its initial state $\ket{E_0}$ to the other state $\ket{E}$ by absorbing a single particle of $\phi$ while at the same time the field $\phi$ undergoes a transition from the one-particle state $\ket{\Psi_0}$ to the vacuum state $\ket{0}$.

By the first-order perturbation theory, the amplitude for the transition
\begin{equation}\label{transition}
\ket{\Psi_0,E_0}\rightarrow\ket{0,E}
\end{equation}
is given by
\begin{eqnarray}\label{A}
\mathcal{A}
&=&
i\kappa\int_{-\infty}^\infty dt \chi(t)
\bra{0,E} \mu(t) \phi(0,t) \ket{\Psi_0,E_0} \nonumber\\
&\equiv&i\kappa\int_{-\infty}^\infty dt \chi(t)
\bra{0,E} \mu(t) \phi(0,t=0) \ket{\Psi(t),E_0}, \qquad
\end{eqnarray}
where the evolution of $\phi$ can be interchangeably described either in the Heisenberg picture or in the Schr\"{o}dinger picture.
Note that the perturbation theory formally requires $\abs{\mathcal{A}}$ to be sufficiently small. If we simply set $\chi(t)=1$, integral over $t\in(-\infty,\infty)$ may yield $\abs{\mathcal{A}}\gtrsim1$ no matter how small $\kappa$ is, thus invalidating the perturbation theory. (See \appref{app:perturbation theory} for the case of a divergent behavior without proper regularization.) Imposing the switching function with a finite support in time (i.e., nonzero only for a finite period of time) can be viewed as a prescription of regularization that makes sense of the perturbation method. The smaller the coupling constant $\kappa$ is, the longer the finite support of $\chi(t)$ can be prescribed for keeping the perturbation method legitimate.

The equation of evolution for $\mu(\tau)$ is given by
\begin{equation}
\mu(t) = e^{iH_0t}\mu(0)e^{-iH_0t},
\end{equation}
where $H_0$ is the Hamiltonian of the detector.
Consequently, we have
\begin{equation}\label{A'}
\mathcal{A} =
i\kappa\bra{E} \mu(0) \ket{E_0}
\int_{-\infty}^\infty dt \chi(t) e^{i(E-E_0)t} \bra{0}\phi(0,0)\ket{\Psi(t)}.
\end{equation}
The transition probability is given by the squared norm of $\mathcal{A}$ as
\begin{eqnarray}\label{P}
&&\mathcal{P} = \mathcal{A}\mathcal{A}^* \nonumber\\
&=& \kappa^2 \abs{\bra{E} \mu(0) \ket{E_0}}^2
\int_{-\infty}^\infty dt \int_{-\infty}^\infty dt'
\chi(t)\chi(t') e^{i\Delta E(t-t')} \nonumber\\
&& \mbox{} \times \bra{0}\phi(0,0)\ket{\Psi(t)} \bra{0}\phi(0,0)\ket{\Psi(t')}^*.
\end{eqnarray}
Substituting \eqref{Psi D1} or \eqref{Psi D2} for $\ket{\Psi(t)}$, we have
\begin{eqnarray}
&&\bra{0}\phi(0,0)\ket{\Psi(t)} \nonumber\\
&=& \frac{1}{2}\int_{-\infty}^\infty dy D(0,0;y,0)
\,e^{-i\delta\omega_0t}
\Big(
f(L_1+y-v_0t) \nonumber\\
&& \qquad\qquad\mbox{} \pm e^{i\theta}f(L_2+y-v_0t)
\Big),
\end{eqnarray}
where
\begin{equation}\label{D}
D(x,t;x',t'):= \bra{0}\phi(x,t)\phi(x',t')\ket{0}
\end{equation}
is the Wightman function.
The probability given by \eqref{P} can be recast as
\begin{eqnarray}\label{P in terms of AL}
\mathcal{P}_{\mathrm{D}_{1,2}} &=& \frac{1}{4}\Big(\mathcal{A}_{L_1}\mathcal{A}_{L_1}^* + \mathcal{A}_{L_2}\mathcal{A}_{L_2}^*
\pm e^{-i\theta}\mathcal{A}_{L_1}\mathcal{A}_{L_2}^* \nonumber\\
&&\quad\mbox{}\pm e^{i\theta}\mathcal{A}_{L_2}\mathcal{A}_{L_1}^*\Big),
\end{eqnarray}
where
\begin{eqnarray}\label{AL}
\mathcal{A}_L &:=&
\kappa \bra{E} \mu(0) \ket{E_0}
\int_{-\infty}^\infty dt
\chi(t) e^{i(\Delta E-\delta\omega_0)t} \nonumber\\
&& \quad \mbox{} \times
\int_{-\infty}^\infty dy D(0,0;y,0) f(L+y-v_0t).\quad
\end{eqnarray}
From now on, we assume that the coupling constant $\kappa$ is small enough to the extent that the finite support of $\chi(t)$ can be prescribed long enough so that it well covers the two arriving wave packages as depicted on the left of \figref{fig:functions}. Accordingly, we can simply set $\chi(\tau)=1$ in \eqref{AL}.
We will discuss what happens if $\kappa$ is large in \secref{sec:remarks} and what happens if the switch-on period of $\chi(t)$ is short in \secref{sec:quantum vs classical}.

Substituting \eqref{phi x t} into \eqref{D} gives the explicit expression for the Wightman function:
\begin{equation}
  D(x,t;x',t')= \int \frac{dk}{2\pi}\frac{1}{2\omega_k} e^{ik(x-x')-i\omega_k(t-t')}.
\end{equation}
Performing the change of variables $y'=L+y-v_0t$ and $t'=t$ upon \eqref{AL}, we obtain (see \appref{app:AL result} for the detailed derivation)
\begin{equation}\label{AL result}
\mathcal{A}_L
= e^{ik_*L} \mathcal{A}_0(k_*),
\end{equation}
where
\begin{equation}\label{k*}
k_* = \frac{1}{v_0}\left(\Delta E - \omega_0 + v_0k_0\right),
\end{equation}
and the $L$-independent part $\mathcal{A}_0(k_*)$ is given by
\begin{eqnarray}\label{A0}
\mathcal{A}_0(k_*) &:=&
\kappa \bra{E} \mu(0) \ket{E_0} \frac{\sqrt{2\pi}\, \tilde{f}(k_*)}{2\omega_{k_*}\abs{g'(k_*)}} \\
&=& \left(\frac{\pi\Delta^2}{2}\right)^{1/4} \frac{\kappa \bra{E} \mu(0) \ket{E_0}}{\omega_{k_*}v_0}
e^{-(k_*-k_0)^2\Delta^2}, \nonumber
\end{eqnarray}
where $g(k)$ is defined in \eqref{g(k)}.
Substituting \eqref{AL result} into \eqref{P in terms of AL}, finally, we obtain the transition probability taking the form
\begin{eqnarray}\label{P result}
\mathcal{P}_{\mathrm{D}_{1,2}}
&=&\frac{1}{2}\big(1 \pm \cos\left(k_*\Delta L+\theta\right)\big)\abs{\mathcal{A}_0(k_*)}^2 \nonumber\\
&=& \sqrt{\frac{\pi\Delta^2}{8}}\,
    \frac{\abs{\kappa \bra{E} \mu(0) \ket{E_0}}^2}{(\omega_{k_*} v_0)^2}\,
     e^{-2(k_*-k_0)^2 \Delta^2}\nonumber\\
&& \mbox{}\times \big(1 \pm \cos\left(k_*\Delta L+\theta\right)\big),
\end{eqnarray}
which is modulated in response to $\theta$.
As the detection probabilities at $\mathrm{D}_1$ and $\mathrm{D}_2$ appear as modulated in response to $\theta$, the wave nature of the interference between $\mathrm{Path_1}$ and $\mathrm{Path_2}$ is manifested.
Unlike the classical counterpart given in \eqref{classical P result}, the modulation pattern of interference does not dims out when $\Delta L$ becomes considerably larger than $\Delta$.

Meanwhile, in order to have $\mathcal{P}_{\mathrm{D}_{1,2}}$ detectable, $k_*$ needs to satisfy
\begin{equation}\label{condition for k*}
\abs{k_*-k_0}\lesssim\frac{\eta}{\Delta},
\end{equation}
where $\eta$ is a numerical factor of $O(1)$, otherwise the factor $e^{-2(k_*-k_0)^2 \Delta^2}$ in \eqref{P result} will render $\mathcal{P}_{\mathrm{D}_{1,2}}$ inappreciable. The condition for obtaining the optimal detection probability is given by $k_*\approx k_0$. By \eqref{k*}, it corresponds to the condition of Fermi's golden rule:
\begin{equation}\label{Fermi's golden rule}
\omega_0 \equiv \sqrt{k_0^2+m^2} \approx \Delta E,
\end{equation}
which tells that the optimal result occurs when the average energy of the incoming single-particle pulse is close to the detector's energy gap. (See \secref{sec:remarks} and \appref{app:perturbation theory} for more discussions on Fermi's golden rule.)

\subsection{Remarks}\label{sec:remarks}
Even when the condition \eqref{Fermi's golden rule} is met, it should be caveated that the result given by \eqref{P result} gives low efficiency of detection far below 100\% (i.e.\ for a single-particle pulse fired from the emitter, it is with high probability that both $\mathrm{D}_1$ and $\mathrm{D}_2$ miss the signal of it), because $\kappa$ is assumed to be sufficiently small. This stands in marked contrast to classical detectors, which in principle can reach nearly perfect efficiency.

One might attempt to increase the coupling constant $\kappa$ to improve the efficiency.
Increasing $\kappa$, however, makes $H_\mathrm{int}$ in \eqref{H int} stronger and consequently increases the energy uncertainty $\delta E\sim\abs{\kappa\bra{E}\mu(0)\ket{E_0}}$ upon the unperturbed energy states $\ket{E_0}$ and $\ket{E}$ of the detector. The uncertainty relation between energy and time imposes a time scale $\delta t\sim1/(2\delta E)\propto 1/\kappa$ upon the detector. This time scale $\delta t$ is what the finite support of $\chi(t)$ has to be prescribed as \eqref{A} is to be regularized.
As $\kappa$ increases, $\delta t$ decreases. If $\delta t$ decreases to the extent that $\delta t\lesssim\Delta L/v_0$ or $\delta t\lesssim\Delta/v_0$, the two arriving wave packages are no longer well covered by the finite support of $\chi(t)$, and consequently our result of the two-path interference given by \eqref{P result} no longer holds correct.
This can also be understood as a consequence of the fact that, as $\kappa$ becomes larger and larger, the quantum detector gradually looses coherence in time as $\delta t$ becomes shorter and shorter.
In the limiting case that $\kappa$ is sufficiently large so that $\delta t\ll\Delta/v_0$, the quantum detector's coherence in time becomes insignificant and it in effect simply behaves like a classical detector.

In reality, the quantum detector's coherence in time, denoted as $\delta t$, is not only determined by $\kappa$ but also diminished by various disturbances in its surroundings that are not included in the perturbation Hamiltonian $H_\mathrm{int}$. Given a finite $\delta t$, the result of \eqref{P result} holds true only if the following conditions are both satisfied:
\begin{subequations}\label{delta t condition}
\begin{eqnarray}
\Delta L &\lesssim& v_0\delta t,\\
\Delta &\lesssim& v_0\delta t,
\end{eqnarray}
\end{subequations}
up to numerical factors of $O(1)$.
That is, the modulation pattern given by \eqref{P result} is the combined result of the interference \emph{between} the two wave packages traveling along the two different pathes as well as that \emph{within} each wave package.

Another obvious attempt to increase the efficiency of detection is to provide, instead of a single quantum detector, an ensemble of identical quantum detectors each of which is coupled to the matter field \emph{independently}. This is possible in principle, but in reality it is extremely difficult to keep the individual detectors decoupled from one another. Any considerable coupling between individual detectors renders the ensemble as a whole incoherent in time and consequently the whole system simply behaves as a classical detector.
In \secref{sec:quantum vs classical}, we will carry out a more formal and rigorous analysis as to how an ensemble of quantum detectors behaves collectively as a classical detector, if each individual detector loses its coherence in time.

Finally, it should be noted that \eqref{P result} is derived particularly based on the \emph{monopole} interaction between the Unruh-DeWitt detector and the scalar field as formulated in \eqref{H int}. For the more realistic case of a two-state quantum detector interacted with photons as an example, the leading effect of the coupling to electromagnetic waves is the magnetic \emph{dipole} (M1) transition (see \appref{app:perturbation theory}). For such a system, the result of \eqref{P result} has to be modified, but it is still expected to yield a two-path interference pattern similar to that in \eqref{P result} provided that the detector's coherence in time is long enough.

\section{Quantum vs.\ classical detectors}\label{sec:quantum vs classical}
The transition probability measured by quantum detectors is given by \eqref{P result}, which is characteristically different from the classical counterpart given by \eqref{classical P result}. As remarked in \secref{sec:remarks} above, the collective result of an ensemble of quantum detectors should reproduce the classical result, if coherence in time of each individual quantum detector becomes sufficiently short.
The underlying mechanism of decoherence in time could be very complicated and difficult to fully understand, but it can be formally modeled by taking the limit that the switching function $\chi(t)$ is turned on very briefly in an ensemble. Modeling $\chi(t)$ with a finite support makes the calculation unmanageable. A tractable treatment, instead, is to model $\chi(t)$ in a Gaussian form as
\begin{equation}\label{chi}
\chi(t)=e^{-(t-t_\chi)^2/2\Delta_\chi^2},
\end{equation}
where $t_\chi$ represents the moment when $\chi(t)$ is fully turned on and $\Delta_\chi$ represents how brief the switch-on is.\footnote{\label{foot:no normalization}Note that the peak value of $\chi(t)$ is assumed to be 1. We should not normalize $\chi(t)$ as $(\sqrt{2\pi}\Delta_\chi)^{-1}e^{-(t-t_\chi)^2/2\Delta_\chi^2}$, which is no longer dimensionless.}
For an ensemble, we assume $\Delta_\chi$ to be constant and model $t_\chi$ as a random variable. In the end, we can take the limit of $v_0\Delta_\chi\ll\Delta$ to reproduce the classical result. On the other hand, the opposite limit of $v_0\Delta_\chi\gg\Delta$ and $v_0\Delta_\chi\gg\Delta L$ is supposed to reproduce the quantum result.
The parameter $\Delta_\chi$ can be viewed as a \emph{formal} prescription for presenting a finite period $\delta t$ of coherence time as mentioned in \secref{sec:remarks}.

The notion of ``ensemble'' here not only describes a true ensemble composed of \emph{many} quantum detectors but can also represent a \emph{single} quantum detector that is in a mixed quantum state corresponding to a probabilistic mixture of random values of $t_\chi$. A quantum detector in such a mixed quantum state is described by the density matrix:
\begin{equation}\label{rho 0}
\rho_0 = \int dt_\chi\, p(t_\chi)\,\ket{E_0^{(\chi(t))}}\bra{E_0^{(\chi(t))}},
\end{equation}
where $\ket{E_0^{(\chi(t))}}$ is the same detector state as appears in \eqref{transition} except that now an additional superscript $(\chi(t))$ is attached to explicitly indicate that this state is associated with a specific switching function, and $p(t_\chi)$ is a probability density function that describes the probabilistic mixture of $t_\chi$.\footnote{More generally, the density state $\rho_0$ can be described as
\begin{equation*}
\rho_0 = \int d\Delta_\chi dt_\chi\, p(\Delta_\chi,t_\chi)\,\ket{E_0^{(\chi(t))}}\bra{E_0^{(\chi(t))}},
\end{equation*}
which exhibits probabilistic mixture not only of $t_\chi$ but also of $\Delta_\chi$. As our purpose is to study how different values of $\Delta_\chi$ give different detection results, we do not consider the mixture of $\Delta_\chi$.}
If the form of \eqref{chi} is prescribed for $\chi(t)$, the density matrix \eqref{rho 0} can be used to model a quantum detector whose coherence in time is characterized by $\Delta_\chi$.\footnote{Consider a pure quantum state as a linear superposition of the two states $\ket{1}$ and $\ket{2}$: $\ket{\psi}=\abs{a}\ket{1}+\abs{b}e^{i\theta}\ket{2}$. The relative phase $e^{i\theta}$ in principle can be measured by an interference experiment.
If, however, the coherence in $\theta$ is corrupted to a certain degree, the pure state $\ket{\psi}$ is reduced into a mixed state: $\rho=\int d\theta\, p(\theta)\ket{\psi}\bra{\psi}$, whereby $\theta$ is smeared with the probability density function $p(\theta)$. In the same spirit, even though we do not know the underlying mechanism of decoherence, the density matrix \eqref{rho 0} provides an adequate formal model for a quantum detector whose coherence in time is corrupted to a certain degree (quantified by $\Delta_\chi$) due to decoherence.}
The initial state of the particle-detector composite system is then given by the density matrix
\begin{eqnarray}
\mathscr{P}_0&\equiv&\ket{\Psi_0}\bra{\Psi_0}\otimes\rho_0 \nonumber\\
&=&\int dt_\chi\, p(t_\chi)\,\ket{\Psi_0,E_0^{(\chi(t))}}\bra{\Psi_0,E_0^{(\chi(t))}},
\end{eqnarray}
where $\ket{\Psi_0}$ is the same single-particle state as appears in \eqref{transition}.
According to the first-order perturbation theory, the state $\ket{\Psi_0,E_0^{(\chi(t))}}$ evolves into
\begin{equation}
\ket{\Psi_0,E_0^{(\chi(t))}} \rightarrow \ket{\Psi_0,E_0^{(\chi(t))}} + \mathcal{A}\,\ket{0,E} + O(\kappa^2),
\end{equation}
where $\mathcal{A}$ is the transition amplitude given by \eqref{A}, which depends on $\chi(t)$. Correspondingly, the matrix density $\mathscr{P}_0$ evolves into
\begin{eqnarray}
\mathscr{P}_0 \rightarrow \mathscr{P} &=& \int dt_\chi\, p(t_\chi) \left(\ket{\Psi_0,E_0^{(\chi(t))}} + \mathcal{A}\,\ket{0,E}\right) \nonumber\\
&&\qquad\mbox{}\times \left(\bra{\Psi_0,E_0^{(\chi(t))}} + \mathcal{A}^*\bra{0,E}\right) \nonumber\\
&& \mbox{} +O(\kappa^3).
\end{eqnarray}
The probability that the quantum detector in the mixed state registers a signal is then given by
\begin{eqnarray}\label{P tilde}
\overline{\mathcal{P}} &=& \mathrm{Tr}\,\big(\ket{0,E}\bra{0,E}\mathscr{P}\big)
=\bra{0,E}\mathscr{P}\ket{0,E} \nonumber\\
&=& \int dt_\chi\, p(t_\chi) \mathcal{A}\mathcal{A}^*
= \int dt_\chi\, p(t_\chi) \mathcal{P},
\end{eqnarray}
where $\mathcal{P}$ is the transition probability given by \eqref{P}.
The final expression of $\overline{\mathcal{P}}$ in \eqref{P tilde} takes the form of an \emph{ensemble average} of $\mathcal{P}$ in the following sense.

Consider an ensemble composed of \emph{many} identical quantum detectors with the same coupling constat $\kappa$ and energy gap $\Delta E$. Suppose each quantum detector is switched on for the same duration $\Delta_\chi$ but at different moments $t_\chi$ that are completely uncorrelated to one another and characterized by the probability density function $p(t_\chi)$.
The average probability (averaged over the ensemble) that the ensemble as a whole registers a signal is obviously given by the same final expression of \eqref{P tilde}.
Therefore, the ensemble average $\overline{\mathcal{P}}$ of $\mathcal{P}$ represents not only the averaged collective behavior of an ensemble of many quantum detectors but also the behavior of a single quantum detector in a mixed state.\footnote{Note that, for a many-detector ensemble, the \emph{total} probability of detection is the \emph{ensemble average} probability multiplied by the size of the ensemble. In principle, by increasing the ensemble size, the detection efficiency of the ensemble as a whole can be made arbitrarily close to 100\% even if $\kappa$ remains small. By contrast, for a single quantum detector in a mixed state, the ``ensemble average'' probability is the total probability, which yields low detection efficiency far below 100\%.}
In most realistic situations, individual quantum detectors in an ensemble are not turned on and off randomly but remains sensitive to the incoming particle all the time. For example, microscopic light-sensitive crystals (as individual quantum detectors) on a photographic film cannot be easily ``turned off''. Nevertheless, the collective behavior of a many-detector ensemble can still be understood in terms of the \emph{formal} average $\overline{\mathcal{P}}$, because each individual quantum detector of the ensemble is in a mixed quantum state as a consequence of decoherence.

To compute the ensemble average, we further model the random variable $t_\chi$ as uniformly distributed in a period of time $t\in[-T,T]$ of our interest (see the right panel of \figref{fig:functions} for illustration). The ensemble average of some quantity $\mathcal{Q}$ is then proportional to $(\Delta_\chi)^{-1}\int_{-T}^T dt_\chi \mathcal{Q}$ up to an overall dimensionless factor whose exact numerical value depends on the formally prescribed form of $\chi(t)$ and is of little interest for our purpose.
Since the exact value of the delimiter $T$ is unimportant as long as the interval $[-T,T]$ is large enough to cover the arrival of the two incoming wave packages, we take the limit $T\rightarrow\infty$ and compute the corresponding ensemble average of $\mathcal{Q}$ up to a proportionality factor as $(\Delta_\chi)^{-1}\int_{-\infty}^\infty dt_\chi \mathcal{Q}$.

As $\mathcal{P}$ is given by \eqref{P in terms of AL}, to compute $\overline{\mathcal{P}}$, we first have to substitute \eqref{chi} into \eqref{AL} to recompute $\mathcal{A}_L$.
Applying various mathematical techniques (see \appref{app:AL result chi} for the detailed derivation), we obtain
\begin{eqnarray}\label{AL result chi}
\mathcal{A}_L &=&
\sqrt{2\pi}\,\Delta_\chi\kappa \bra{E} \mu(0) \ket{E_0}
\int_{-\infty}^\infty dz \int_{-\infty}^\infty\frac{dk}{2\pi}\frac{1}{2\omega_k} \nonumber\\
&&\quad\mbox{}\times e^{ig(k)t_\chi-8\Delta_\chi^2g(k)^2} e^{-ik(z-L)} f(z),
\end{eqnarray}
where $g(k)$ is defined in \eqref{g(k)}.
Consequently, the corresponding ensemble average of $\mathcal{A}_{L_1} \mathcal{A}^*_{L_2}$ up to a proportionality factor is given by (see \appref{app:deriving average AAstar} for the detailed derivation)
\begin{eqnarray}\label{deriving average AAstar}
&& \overline{\mathcal{A}_{L_1} \mathcal{A}^*_{L_2}} \propto \frac{1}{\Delta_\chi}\int_{-\infty}^\infty dt_\chi \mathcal{A}_{L_1} \mathcal{A}^*_{L_2} \nonumber\\
&=&\sqrt{\frac{\pi}{2}}\,\Delta\Delta_\chi\abs{\kappa \bra{E} \mu(0) \ket{E_0}}^2
\int_{-\infty}^\infty dk\frac{e^{-h(k)-ik\Delta L}}{\omega_k^2\abs{g'(k)}},\qquad
\end{eqnarray}
where $h(k)$ is defined as
\begin{eqnarray}\label{def h}
h(k) &:=& 16\Delta_\chi^2 g(k)^2 + 2\Delta^2 (k-k_0)^2 \nonumber\\
&=& 16\Delta_\chi^2v_0^2(k-k_*)^2 + 2\Delta^2 (k-k_0)^2,
\end{eqnarray}
The factor $e^{-h(k)-ik\Delta L}$ appearing in the final line of \eqref{deriving average AAstar} is a Gaussian package over $k$ centered at
\begin{equation}\label{k medstar}
k_\medstar := \frac{8v_0^2\Delta_\chi^2k_*+\Delta^2k_0}{8v_0^2\Delta_\chi^2+\Delta^2}
\end{equation}
with the variance given by $(8v_0^2\Delta_\chi^2+\Delta^2)^{-1}< \Delta^{-2}$.
On the grounds that $e^{-h(k)-ik\Delta L}$ is sharply centered at $k=k_\medstar$ with variance smaller than $\Delta^{-2}$, it is a good approximation to move the factor $(\omega_k^2\abs{g'(k)})^{-1}$ out of the integral by setting $k=k_\medstar$.
That is, we have
\begin{eqnarray}\label{ensemble average AA*}
& &\frac{1}{\Delta_\chi}\int_{-\infty}^\infty dt_\chi \mathcal{A}_{L_1} \mathcal{A}^*_{L_2} \nonumber\\
&\approx& \sqrt{\frac{\pi}{2}}\frac{\Delta\Delta_\chi\abs{\kappa \bra{E} \mu(0) \ket{E_0}}^2}{\omega_{k_\medstar}^2\abs{g'(k_\medstar)}}
\int_{-\infty}^\infty dk\, e^{-h(k)-ik\Delta L} \nonumber\\
&=& \frac{\pi\Delta}{2} \frac{\abs{\kappa \bra{E} \mu(0) \ket{E_0}}^2}{\omega_{k_\medstar}^2 v_0}\,
          \frac{\Delta_\chi}{ \sqrt{\Delta^2+8v_0^2 \Delta_\chi^2}} \nonumber\\
&& \mbox{}\times \exp\left[-\frac{\Delta L(\Delta L+8ik_0\Delta^2)}{8(\Delta^2+8v_0^2 \Delta_\chi^2)}\right]\nonumber\\
&& \mbox{}\times \exp\left[-\frac{8v_0^2\Delta_\chi^2 \left(2(k_0-k_*)^2\Delta^2+i k_*\Delta L\right)}{\Delta^2+8v_0^2 \Delta_\chi^2}\right],\qquad
\end{eqnarray}
where \eqref{Gaussian formula} has been used.

We first consider the situation when $\Delta_\chi$ is much larger than $\Delta$ and $\Delta L$, or more precisely $v_0\Delta_\chi\gg\Delta$ and $v_0\Delta\chi\gg\Delta L$.
It follows from \eqref{k medstar} and \eqref{ensemble average AA*} that
\begin{eqnarray}\label{ensemble average AA* limit infty}
&&\lim_{v_0\Delta_\chi\gg \Delta, \Delta L}\frac{1}{\Delta_\chi}\int_{-\infty}^\infty dt_\chi \mathcal{A}_{L_1} \mathcal{A}^*_{L_2} \nonumber\\
&=&
\frac{\pi\Delta}{4\sqrt{2}}
    \frac{\abs{\kappa \bra{E} \mu(0) \ket{E_0}}^2}{(\omega_{k_*} v_0)^2}\,
    e^{-2(k_0-k_*)^2\Delta^2} e^{-ik_*\Delta L} \nonumber\\
%-------
&=& \frac{\sqrt{\pi}}{4} e^{-ik_*\Delta L}\, \abs{\mathcal{A}_{0}(k_*)}^2,
\end{eqnarray}
where $\mathcal{A}_{0}(k_*)$ is defined in \eqref{A0}.
The transition probability measured as the ensemble average is obtained by replacing each term $\mathcal{A}_{L_1} \mathcal{A}^*_{L_2}$ in \eqref{P in terms of AL} with the ensemble average counterpart \eqref{ensemble average AA* limit infty}. The result takes the form
\begin{equation}\label{P result as a limit}
\overline{\mathcal{P}}_{\mathrm{D}_{1,2}}^{(v_0\Delta_\chi\gg \Delta, \Delta L)} \propto \frac{1}{2}\big(1 \pm \cos\left(k_*\Delta L+\theta\right)\big)\abs{\mathcal{A}_0(k_*)}^2,
\end{equation}
which is identical to \eqref{P result} up to a proportionality factor $\sqrt{\pi}/4$.\footnote{The factor $\sqrt{\pi}/4$ arises as a consequence of modeling $\chi(t)$ by a Gaussian function with width $\Delta_\chi$ as given in \eqref{chi}, instead of a rectangular function with width $\Delta_\chi$.}
We have shown that the transition probability measured as the ensemble average is characteristically identical to that of a single quantum detector if each detector of the ensemble remains perfectly coherent in time.

Next, we consider the other extreme that $\Delta_\chi$ is much smaller than $\Delta$, or more precisely $v_0\Delta_\chi\ll\Delta$. Again, by \eqref{k medstar} and \eqref{ensemble average AA*}, we have
\begin{eqnarray}\label{classical AA* limit}
&& \lim_{v_0\Delta_\chi\ll\Delta}\frac{1}{\Delta_\chi}\int_{-\infty}^\infty dt_\chi \mathcal{A}_{L_1} \mathcal{A}^*_{L_2} \nonumber\\
&=& \frac{\pi}{2}\Delta_\chi e^{-16v_0^2\Delta_\chi^2(k_0-k_*)^2} \nonumber\\
&& \quad \mbox{}\times
\frac{\abs{\kappa \bra{E} \mu(0) \ket{E_0}}^2}{\omega_{k_0}^2 v_0}
e^{-ik_0 \Delta L}\,
e^{-(\frac{\Delta L}{2\sqrt{2}\Delta})^2}.
\end{eqnarray}
Finally, replacing each term $\mathcal{A}_{L_1} \mathcal{A}^*_{L_2}$ in \eqref{P in terms of AL} with the correspondingly counterpart \eqref{classical AA* limit}, we obtain
\begin{equation}\label{reduced to classical}
\overline{\mathcal{P}}_{\mathrm{D}_{1,2}}^{(v_0\Delta_\chi\ll\Delta)} \propto \frac{1}{2}\left(1\pm\cos(k_0\Delta L+\theta)\,
e^{-\left(\frac{\Delta L}{2\sqrt{2}\Delta}\right)^2} \right),
\end{equation}
which is identical to \eqref{classical P result} up to a proportionality constant, which accounts for the detection efficiency.
Therefore, we have formally shown that an ensemble of identical quantum detectors as a whole indeed behave as a classical detector, if the coherence in time of each individual detector becomes sufficiently shorter than $\Delta$.

It is noteworthy that the overall proportionality factor in \eqref{reduced to classical} contains a notable factor in a Gaussian form as
\begin{equation}\label{proportional factor}
\overline{\mathcal{P}}_{\mathrm{D}_{1,2}}^{(v_0\Delta_\chi\ll\Delta)} \propto \Delta_\chi e^{-16v_0^2\Delta_\chi^2(k_0-k_*)^2}.
\end{equation}
If $v_0\Delta_\chi\ll\lambda_0\equiv2\pi/k_0\ll\Delta$, this factor simply flattens into $\Delta_\chi$. It tells that the energy gap $\Delta E\equiv\omega_0+v_0(k_*-k_0)$ shows no particular preference for $k_0$.
On the other hand, if $\lambda_0\equiv2\pi/k_0\ll v_0\Delta_\chi\ll\Delta$, this factor is picked at $k_0=k_*$. It tells that $\Delta E$ favors $k_0$ that is close to $k_*$, or equivalently $\omega_0\approx\Delta E$.
These results agree with the familiar features pertaining to Fermi's golden rule (see \secref{sec:remarks 2} and \appref{app:perturbation theory}, especially the text after \eqref{Delta E and omega}, for more discussions).

The ensemble average $\overline{\mathcal{P}}$ as calculated above represents not only the averaged collective detection probability of a large ensemble of many identical quantum detectors but also the detection probability of a single quantum detector in a mixed quantum state.
For a single quantum detector, if we manage to corrupt its coherence in time (while keeping it turned on all the time), it will be in a mixed state described by \eqref{rho 0} instead of a pure state. Consequently, in regard to the accumulated count of individual signals, it behaves effectively like a classical detector (but with very low detection efficiency). This can be achieved, for example, by coupling the quantum detector to a high-precision clock that measures occurrence time of the signal. Therefore, for a quantum detector to behave as a genuine Unruh-DeWitt-type detector, it has to remain coherent in time and thus agnostic of the signal occurrence time.\footnote{As remarked in Footnote \footref{foot:distinction}, the difference between the classical and quantum behaviors essentially relies on whether a detector can remain \emph{coherent} in time, instead of whether it remains \emph{ignorant} of time. The latter is a necessary condition for the former, but not sufficient.}

In the analysis presented above, we do not attempt to formulate the underlying mechanism of quantum decoherence, which may result from various complicated processes as remarked in \secref{sec:remarks}. Instead, we study what will happen if the coherence in time is lost by formally reducing the time span of the switching function $\chi(t)$ for each Unruh-DeWitt detector in an ensemble.
The formal result nevertheless substantiates the main idea of \emph{decoherence theory} (see \cite{schlosshauer2005decoherence} for a review): \emph{quantum behavior is lost as a result of quantum decoherence.}\footnote{Decoherence theory provides an explanation for quantum collapse, but it has to be noted that it has not solved the measurement problem. See \cite{adler2003decoherence} for the comments.}

As our model yields a substantial difference between \eqref{classical P result} for ordinary classical detectors and \eqref{P result} for Unruh-DeWitt-type quantum detectors, and shows that the former can be viewed as a formal limit of the latter, it might provide new insight into the measurement problem in quantum mechanics.
Particularly, the result of our formal model seems to support the main idea of \emph{objective-collapse theories} (e.g.\ \cite{ghirardi1985model, ghirardi1986unified, ghirardi1990markov, penrose1996gravity, penrose1998quantum, penrose2014gravitization, jabs2016conjecture}) that a quantum state in superposition is collapsed into a definite state when a certain objective physical threshold is reached (e.g., $v_0\Delta_\chi\ll\Delta$ in our model).

\subsection{Remarks}\label{sec:remarks 2}
It should be emphasized that the result measured by a quantum detector as given in \eqref{P result} or \eqref{P result as a limit} is markedly different from that measured by a classical detector with an etalon (i.e., Fabry-P\'{e}rot resonator) placed in front of it. An etalon filters most frequency modes of the incoming pulse and allows nearly only one single mode at the etalon's resonant frequency $\omega_q\equiv(k_q^2+m^2)^{1/2}$ to enter the detector. Provided that the etalon has high finesse, it effectively changes the incoming pulse into a nearly monochromatic plane wave. In other words, the profile $\tilde{f}(k)$ in \eqref{f tilde} is altered by the etalon: $k_0$ is replaced by $k_q$ and $\Delta$ is replaced by $\Delta_q$, the inverse of the bandwidth of the etalon. If the finesse is high enough so that $\Delta_q\gg\Delta L$, the detection probability of a classical detector with the etalon in front of it, according to \eqref{classical P result}, takes the form
\begin{eqnarray}\label{P with etalon}
\mathcal{P}_{\mathrm{D}_{1,2}}^\mathrm{(cl+etalon)}
&\propto&
\frac{1}{2}\left(1\pm\cos(k_q\Delta L+\theta)e^{-\left(\frac{\Delta L}{2\sqrt{2}\Delta_q}\right)^2}\right) \nonumber\\
&\approx&
\frac{1}{2}\left(1\pm\cos(k_q\Delta L+\theta)\right),
\end{eqnarray}
where the proportionality constant, i.e.\ the efficiency of detection, is inevitably reduced by the etalon, and its  exact value depends on the etalon's finesse and how much $k_q$ differs from $k_0$. The result of \eqref{P with etalon} apparently restores the interference pattern, even if $\Delta L$ is larger than the original $\Delta$. However, the restoration is achieved by altering the waveform of the original pulse, and the properties of the original parameters $k_0$ and $\Delta$ are obliterated in \eqref{P with etalon}. We do not say that the two-path interference pattern is manifested even if $\Delta L$ considerably exceeds the coherence length of the single-particle pulse, for the coherence length of the pulse is in fact greatly prolonged by the etalon.
By contrast, the result measured by a quantum detector as given in \eqref{P result} manifests the two-path interference pattern without altering the incoming pulse at all, and it indeed depends on both $k_0$ and $\Delta$. By comparison, the interference pattern of \eqref{P result} is achieved by prolonging the coherence time \emph{of the detector}, whereas that of \eqref{P with etalon} is achieved by prolonging the coherence time \emph{of the incoming pulse}. The difference is not just a matter of interpretation; rather, \eqref{P result} and \eqref{P with etalon} give two qualitatively different interference patterns.\footnote{If $\Delta$ is small enough, \eqref{condition for k*} tells that $k_*$ can be fairly different from $k_0$ (i.e., $\omega_0$ can be fairly different from $\Delta E$) to still yield appreciable $\mathcal{P}_{\mathrm{D}_{1,2}}$ in \eqref{P result}, although the optimal result occurs at $\omega_0\approx\Delta E$. By contrast, in \eqref{P with etalon}, regardless of $\Delta$, the etalon's high finesse renders the detection efficiency inappreciable unless $\omega_0\approx\omega_q$ (i.e., $\omega_0$ has to match $\omega_q$ very closely). The result of \eqref{P result} involves the interference between \emph{all} frequency modes of the incoming pulse, whereas the result of \eqref{P with etalon} involves nearly only a \emph{single} frequency mode of $\omega_q$. If one can manage to have an ensemble of identical quantum detectors each of which is coupled to the matter field \emph{independently} (although this is extremely difficult as remarked in \secref{sec:remarks}), the ensemble as a whole will give the result of \eqref{P result as a limit}, which takes the same form of \eqref{P result} but in principle can yield arbitrarily high efficiency by increasing the ensemble size. By contrast, the etalon always filters out some photons and thus the detection efficiency is always considerably lower than 100\%.}

Another attempt to reproduce the two-path interference pattern by a classical detector even if $\Delta L$ considerably exceeds the coherence length $\Delta$ of the single-particle pulse is to make the interaction between the arriving pulses and the detector sufficiently weak (i.e., to weaken the coupling constant $\kappa$ sufficiently), so that the detection bandwidth is sharply narrowed to a single frequency mode that nearly perfectly matches $\Delta E= E-E_0$ according to Fermi's golden rule. As the detector effectively responses only to the single mode, the detection probability takes the form
\begin{equation}\label{P flawed}
\mathcal{P}_{\mathrm{D}_{1,2}}
\propto
\frac{1}{2}\left(1\pm\cos(k_{\Delta E}\Delta L+\theta)\right), \quad \text{(wrong!)}
\end{equation}
where $k_{\Delta E}$ is defined via $\Delta E =: (k_{\Delta E}^2+m^2)^{1/2}$. Although the efficiency of detection might be low due to weak coupling, the detection probability manifests the two-path interference regardless of whether $\Delta L$ exceeds $\Delta$ or not.
However, the above reasoning is flawed, and \eqref{P flawed} turns out to be incorrect.

As remarked in \secref{sec:remarks}, weakening the coupling constant $\kappa$ indeed is a necessary condition to have the detector's coherence in time $\delta t$ long enough, but $\delta t$ is also degraded by various dissipative disturbances from its surroundings. Modeling the detector's coherence in time by the switching function $\chi(t)$ in the form of \eqref{chi}, we can calculate the (ensemble averaged) detection probability as given in \eqref{ensemble average AA*} for arbitrary values of $\delta t \sim \Delta_\chi$.
If one can manage to make $\delta t\sim\Delta_\chi$ long enough by sufficiently isolating the detector from environmental disturbances, the detector simply behaves as a quantum detector (by our terminology), and its detection probability takes the form \eqref{P result as a limit}. Note that the interference pattern in \eqref{P result as a limit} is in response to $k_*$, whereas the pattern in \eqref{P flawed} is to $k_{\Delta E}$.
If, on the other hand, $\delta t \sim \Delta_\chi$ is sufficiently short (despite the fact that $\kappa$ is very small), the detector simply behaves as a classical detector, and its detection probability takes the form \eqref{reduced to classical}, which renders the interference pattern disappeared when $\Delta L\gg\Delta$ and is qualitatively very different from \eqref{P flawed}.

In any case, \eqref{ensemble average AA*} does not give rise to the form of \eqref{P flawed} whatever value $\delta t\sim\Delta_\chi$ is taken, and in fact \eqref{P flawed} is fallacious. The fallacy of the reasoning towards \eqref{P flawed} is because the conventional wisdom of Fermi's golden rule, which is often taken for granted, does not apply when the interference between different frequency modes become relevant.
See \appref{app:perturbation theory} for more elaborations on when and why Fermi's golden rule requires careful reconsideration.
As commented after \eqref{proportional factor}, in the classical limit that $v_0\Delta_\chi\ll\lambda_0\equiv2\pi/k_0\ll\Delta$, the energy gap $\Delta E$ shows no particular preference for $k_0$. Meanwhile, in the classical limit that $\lambda_0\equiv2\pi/k_0\ll v_0\Delta_\chi\ll\Delta$, the detection probability is optimal when $\omega_0\approx\Delta E$, but remains nonzero even if $\omega_0$ is deviated from $\Delta E$. On the other hand, as commented after \eqref{Fermi's golden rule}, in the quantum limit that $v_0\Delta_\chi\gg\Delta$ and $v_0\Delta\chi\gg\Delta L$, the detection probability again is optimal when $\omega_0\approx\Delta E$, but remains nonzero otherwise. This shows that Fermi's golden rule in the strict form (i.e., formulated as a delta function selection rule) can be largely soften or even broken when the interference between different frequency modes are taken into account.\footnote{One might wonder whether it is problematic if the strict form of Fermi's golden rule does not hold, as conservation of energy seems to be violated. The answer is negative. The apparent energy mismatch between the initial and final states is supposed to be transferred into or from other energy forms that are not taken into account in our calculation, such as the center-of-mass kinetic energy of the detector or the energy of the surroundings that mount the detector.}

To sum up, the detection probability measured by a quantum detector as given in \eqref{P result} or \eqref{P result as a limit} manifesting the two-path interference pattern even if $\Delta L$ considerably exceeds $\Delta$ is a genuinely novel result that cannot be reproduced using a classical detector and has not been observed yet. In \secref{sec:discussion}, it will be commented that the technology required to produce the quantum result is extremely challenging.

\section{The quantum system of one particle and two detectors}\label{sec:two detectors}
In the analysis of \secref{sec:quantum detectors}, we have assumed that the two quantum detectors $\mathrm{D}_1$ and $\mathrm{D}_2$ can be treated separately in the sense that we can focus solely on one of them regardless of the presence or absence of the other. Since $\mathrm{D}_1$ and $\mathrm{D}_2$ are quantum states as well as the particle, we shall, more rigorously, regard the particle and the two detectors altogether as a whole quantum system subject to additional classical detectors that are used to perform the final measurement of the resulting states of $\mathrm{D}_1$ and $\mathrm{D}_2$. In this section, we will show that this rigorous treatment yields the same result as that obtained by treating $\mathrm{D}_1$ and $\mathrm{D}_2$ separately.

When the two quantum detectors are taken into account together, instead of considering \eqref{Psi D1} or \eqref{Psi D2} separately, the wave function of the single-particle pulse is taken to be\footnote{Note that we do not add an overall factor of $1/\sqrt{2}$ here, because the factors of $1/\sqrt{2}$ accounting for the splitting of $\mathrm{BS}_\mathrm{in}$ and $\mathrm{BS}_\mathrm{out}$ have been taken into account as explained in the text after \eqref{F D1'}.}
\begin{eqnarray}
\ket{\Psi(t)} &=& \ket{\Psi^{\mathrm{D}_1}(t)}+\ket{\Psi^{\mathrm{D}_2}(t)} \nonumber\\
&\approx& e^{-i\delta\omega_0t} \int dy F_{\mathrm{D}_1}(y,t)
\phi(y)\ket{0} \nonumber\\
&& \mbox{}+e^{-i\delta\omega_0t}\int dz F_{\mathrm{D}_2}(z,t)
\phi(z)\ket{0},
\end{eqnarray}
where $y$ and $z$ represent the relative locations with reference to $\mathrm{D}_1$ and $\mathrm{D}_2$, respectively, and $F_{\mathrm{D}_1}$ and $F_{\mathrm{D}_2}$ are given by \eqref{F D1} and \eqref{F D2}, respectively.
Now, consider the process that the field $\phi$ undergoes a transition from the one-particle state $\ket{\Psi(t)}$ to the vacuum state $\ket{0}$ and at the same time one of the two detectors is excited from its ground state $\ket{E_0^{\mathrm{D}_1}}$ or $\ket{E_0^{\mathrm{D}_2}}$ to its excited state $\ket{E^{\mathrm{D}_1}}$ or $\ket{E^{\mathrm{D}_2}}$.
In view of the whole system, the interaction Hamiltonian \eqref{H int} is recast as
\begin{eqnarray}
H_\mathrm{int} &=& H^{\mathrm{D}_1}_\mathrm{int} + H^{\mathrm{D}_2}_\mathrm{int} \nonumber\\
&\equiv& \kappa_{\mathrm{D}_1} \chi_{\mathrm{D}_1}(\tau)\mu_{\mathrm{D}_1}(\tau)\phi(x_{\mathrm{D}_1}^\mu(\tau)) \nonumber\\
&& \quad \mbox{} +
\kappa_{\mathrm{D}_2} \chi_{\mathrm{D}_2}(\tau)\mu_{\mathrm{D}_2}(\tau)\phi(x_{\mathrm{D}_2}^\mu(\tau)),
\end{eqnarray}
where $x_{\mathrm{D}_1}^\mu(\tau) = (y=0,t=\tau)$ and $x_{\mathrm{D}_2}^\mu(\tau) = (z=0,t=\tau)$, and we keep it generic that the two quantum detectors may have different monopole moments $\mu_{\mathrm{D}_{1,2}}(t)$, coupling constants $\kappa_{\mathrm{D}_{1,2}}$ and switching functions $\chi_{\mathrm{D}_{1,2}}(t)$.

Prior to the final measurement using classical detectors, the state of the whole system after the interaction between and incoming pulse and the two quantum detectors is changed from $\ket{\Psi(t)}\otimes\ket{E_0^{\mathrm{D}_1},E_0^{\mathrm{D}_2}}$ to a superposition state as
\begin{eqnarray}\label{Psi sys}
\ket{\Psi_\mathrm{sys}(t)} &=&
\ket{0}\otimes\left(\alpha\ket{E^{\mathrm{D}_1},E_0^{\mathrm{D}_2}}+\beta\ket{E_0^{\mathrm{D}_1},E^{\mathrm{D}_2}}\right) \nonumber\\
&& \qquad \quad \mbox{} + \gamma \ket{\Psi(t)}\otimes\ket{E_0^{\mathrm{D}_1},E_0^{\mathrm{D}_2}},
\end{eqnarray}
where $t$ is given as a large number whose exact numerical value is unimportant, and the coefficients $\alpha$, $\beta$, and $\gamma$ are to be computed.\footnote{As we are considering the transition amplitude up to the first-order perturbation, the probability that both $\mathrm{D}_1$ and $\mathrm{D}_2$ register signals is assumed to be extremely negligible.}
Calculating the first-order transition amplitudes from $\ket{\Psi(t)}\otimes\ket{E_0^{\mathrm{D}_1},E_0^{\mathrm{D}_2}}$ into $\ket{0,E^{\mathrm{D}_1},E_0^{\mathrm{D}_2}}$ and $\ket{0,E_0^{\mathrm{D}_1},E^{\mathrm{D}_2}}$ respectively by the same method for \eqref{A} and \eqref{A'}, we obtain
\begin{subequations}\label{alpha and beta}
\begin{eqnarray}
\alpha &=& i\int_{-\infty}^\infty dt
\bra{0,E^{\mathrm{D}_1},E_0^{\mathrm{D}_2}} H_\mathrm{int} \ket{\Psi(t=0),E_0^{\mathrm{D}_1},E_0^{\mathrm{D}_2}} \nonumber\\
&\approx& i\int_{-\infty}^\infty dt
\bra{0,E^{\mathrm{D}_1}} H^{\mathrm{D}_1}_\mathrm{int} \ket{\Psi^{\mathrm{D}_1}(t=0),E_0^{\mathrm{D}_1}}
\nonumber\\
&=& i\kappa_{\mathrm{D}_1}\bra{E^{\mathrm{D}_1}} \mu_{\mathrm{D}_1}(0) \ket{E_0^{\mathrm{D}_1}}
\int_{-\infty}^\infty dt\, \chi_{\mathrm{D}_1}(t) \nonumber\\
&& \mbox{}\times
  e^{i(E^{\mathrm{D}_1}-E_0^{\mathrm{D}_1})t} \bra{0}\phi(y=0,t=0)\ket{\Psi^{\mathrm{D}_1}(t)} \nonumber\\
&\equiv& \mathcal{A}^{\mathrm{D}_1}, \\
%---------------------------------------------------------------
\beta &=& i\int_{-\infty}^\infty dt
\bra{0,E_0^{\mathrm{D}_1},E^{\mathrm{D}_2}} H_\mathrm{int} \ket{\Psi(t=0),E_0^{\mathrm{D}_1},E_0^{\mathrm{D}_2}} \nonumber\\
&\approx& i\int_{-\infty}^\infty dt
\bra{0,E^{\mathrm{D}_2}} H^{\mathrm{D}_2}_\mathrm{int} \ket{\Psi^{\mathrm{D}_2}(t=0),E_0^{\mathrm{D}_2}}
\nonumber\\
&=& i\kappa_{\mathrm{D}_2}\bra{E^{\mathrm{D}_2}} \mu_{\mathrm{D}_2}(0) \ket{E_0^{\mathrm{D}_2}}
\int_{-\infty}^\infty dt\, \chi_{\mathrm{D}_2}(t)  \nonumber\\
&& \mbox{} \times
e^{i(E^{\mathrm{D}_2}-E_0^{\mathrm{D}_2})t} \bra{0}\phi(z=0,t=0)\ket{\Psi^{\mathrm{D}_2}(t)} \nonumber\\
&\equiv& \mathcal{A}^{\mathrm{D}_2},
\end{eqnarray}
\end{subequations}
where we have made the very accurate approximation $H^{\mathrm{D}_1}_\mathrm{int} \ket{\Psi^{\mathrm{D}_2}(t)}\approx0$ and $H^{\mathrm{D}_2}_\mathrm{int} \ket{\Psi^{\mathrm{D}_1}(t)}\approx0$ on the grounds that $\mathrm{D}_1$ is far away from the trajectory of $\ket{\Psi^{\mathrm{D}_2}(t)}$ and $\mathrm{D}_2$ from that of $\ket{\Psi^{\mathrm{D}_1}(t)}$. Also note that $\mathcal{A}^{\mathrm{D}_{1,2}}$ are identical to \eqref{A'} except that various superscripts and subscripts are now specified by $\mathrm{D}_1$ or $\mathrm{D}_2$ for clarity.

Therefore, \eqref{Psi sys} takes the form
\begin{eqnarray}\label{Psi sys result}
\ket{\Psi_\mathrm{sys}(t)} &=&
\ket{0}\otimes\left(\mathcal{A}^{\mathrm{D}_1}\ket{E^{\mathrm{D}_1},E_0^{\mathrm{D}_2}} +\mathcal{A}^{\mathrm{D}_2}\ket{E_0^{\mathrm{D}_1},E^{\mathrm{D}_2}}\right) \nonumber\\
&&\quad \mbox{} + \sqrt{1-\mathcal{P}_{\mathrm{D}_1} -\mathcal{P}_{\mathrm{D}_2}} \,e^{-i(E_0^{\mathrm{D}_1}+E_0^{\mathrm{D}_1})t}\nonumber\\
&& \qquad\quad \mbox{} \times \ket{\Psi(t)}\otimes\ket{E_0^{\mathrm{D}_1},E_0^{\mathrm{D}_2}}, \qquad
\end{eqnarray}
where $\mathcal{P}_{\mathrm{D}_{1,2}}:=\mathcal{A}^{\mathrm{D}_{1,2}}\mathcal{A}^{\mathrm{D}_{1,2}*}$ are the same as \eqref{P} and \eqref{P in terms of AL}.
When the final classical measurement of the statuses of $\mathrm{D}_1$ and $\mathrm{D}_2$ is performed upon $\ket{\Psi_\mathrm{sys}}$, we will have the three \emph{mutually exclusive} outcomes:
\begin{enumerate}[(i)]
\item $\mathrm{D}_1$ registers a signal; i.e., $\mathrm{D}_1$ is in the state $\ket{E^{\mathrm{D}_1}}$.
\item $\mathrm{D}_2$ registers a signal; i.e., $\mathrm{D}_2$ is in the state $\ket{E^{\mathrm{D}_2}}$.
\item Neither $\mathrm{D}_1$ nor $\mathrm{D}_1$ registers a signal; i.e., $\mathrm{D}_1$ remains in the state $\ket{E_0^{\mathrm{D}_1}}$ and $\mathrm{D}_2$ in $\ket{E_0^{\mathrm{D}_2}}$.
\end{enumerate}
These outcomes of the final classical measurement can be neatly associated with the three POVM (positive operator-valued measure) operators:
\begin{subequations}
\begin{eqnarray}
E_\mathrm{(i)} &=& \ket{E^{\mathrm{D}_1}}\bra{E^{\mathrm{D}_1}} \otimes \mathbbm{1}^{\phi\otimes\mathrm{D}_2}, \\
E_\mathrm{(ii)} &=& \ket{E^{\mathrm{D}_2}}\bra{E^{\mathrm{D}_2}} \otimes \mathbbm{1}^{\phi\otimes\mathrm{D}_1}, \\
E_\mathrm{(iii)} &=& \mathbbm{1}^{\phi\otimes\mathrm{D}_1\otimes\mathrm{D}_2} - E_\mathrm{(i)} - E_\mathrm{(ii)}.
\end{eqnarray}
\end{subequations}
The probabilities of these three outcomes are given by $\bra{\Psi_\mathrm{sys}}E_{(a)}\ket{\Psi_\mathrm{sys}}=\mathcal{P}_{(a)}$, which read as $\mathcal{P}_\mathrm{(i)}=\mathcal{P}_{\mathrm{D}_1}$, $\mathcal{P}_\mathrm{(ii)}=\mathcal{P}_{\mathrm{D}_2}$, and $\mathcal{P}_\mathrm{(iii)}=1-\mathcal{P}_{\mathrm{D}_1}-\mathcal{P}_{\mathrm{D}_1}$.

As $\mathcal{P}_{\mathrm{D}_{1,2}}$ is identical to that given in \eqref{P} and \eqref{P in terms of AL}, we have shown that the final classical measurements upon $\mathrm{D}_1$ and $\mathrm{D}_2$ yield the same results as we treat $\mathrm{D}_1$ and $\mathrm{D}_2$ separately.
More precisely, one can focus solely on the final classical measurement upon $\mathrm{D}_1$ while completely disregarding the presence or absence of $\mathrm{D}_2$. The associated POVM operators for the outcomes that $\mathrm{D}_1$ registers a signal and that $\mathrm{D}_1$ does not register a signal are given by
\begin{subequations}
\begin{eqnarray}
E_{(\mathrm{D}_1,+)} &\equiv& E_\mathrm{(i)} = \ket{E^{\mathrm{D}_1}}\bra{E^{\mathrm{D}_1}} \otimes \mathbbm{1}^{\phi\otimes\mathrm{D}_2}, \\
E_{(\mathrm{D}_1,-)} &=& \mathbbm{1}^{\phi\otimes\mathrm{D}_1\otimes\mathrm{D}_2} - E_{(\mathrm{D}_1,+)},
\end{eqnarray}
\end{subequations}
and the corresponding probabilities are given by $\bra{\Psi_\mathrm{sys}}E_{(\mathrm{D}_1,+)}\ket{\Psi_\mathrm{sys}}=\mathcal{P}_{\mathrm{D}_1}$ and $\bra{\Psi_\mathrm{sys}}E_{(\mathrm{D}_1,-)}\ket{\Psi_\mathrm{sys}}=1-\mathcal{P}_{\mathrm{D}_1}$.
This affirms our assumption that, as far as the final classical measurement of the status of $\mathrm{D}_1$ is concerned, the result obtained by considering the quantum system of the particle plus $\mathrm{D}_1$ and $\mathrm{D}_2$ is the same as that obtained by considering the quantum system of the particle plus $\mathrm{D}_1$ only. The same is true if one focuses solely on the final classical measurement upon $\mathrm{D}_2$ while completely disregarding the presence or absence of $\mathrm{D}_1$.

Also note that we have kept the switching function $\chi_{\mathrm{D}_{1,2}}$ generic in \eqref{alpha and beta}. By taking the form of \eqref{chi} and considering the $v_0\Delta_\chi\gg\Delta,\Delta_L$ and $v_0\Delta_\chi\ll\Delta$ limits, respectively, we can produce the results for both classical and quantum detectors as shown in \secref{sec:quantum vs classical}. Therefore, whether $\mathrm{D}_1$ and $\mathrm{D}_2$ are classical or quantum detectors, we can always focus solely on one of them without taking into account the presence or absence of the other.

\section{Summary and discussion}\label{sec:discussion}
We have arrived at the conclusion that, provided the Unruh-DeWitt-type quantum detectors remain coherent in time during the period of the arrival of the two wave packages traveling along the two paths, the detection probability measured by the Unruh-DeWitt detectors is given by \eqref{P result}, which manifests the two-path interference as modulated in response to $\theta$, even if the length difference $\Delta L$ between the two paths considerably exceeds the coherence length $l_c\sim\Delta$ of the single-particle pulse.
By contrast, if measured by ordinary classical detectors, the detection probability is given by \eqref{classical P result}, which contains an exponential diminishing factor that renders the interference pattern invisible once $\Delta L$ exceeds ${\sim}6\Delta$.
The reason for the difference essentially lies in the fact that ordinary classical detectors have no or little coherence in time, whereas the Unruh-DeWitt detectors are assumed to remain coherent long enough and manifest quantum interference \emph{over time}.\footnote{Although it is not often emphasized, the implicit assumption of long coherence in time is essential for deriving the celebrated Unruh effect. See, e.g., the comment on ``the quantum interference over time'' in \cite{chiou2018response}.}

The quantum detector's coherence in time is delimited by its coupling strength with the matter field $\phi$ and can be further degraded by various dissipative interactions with its environment, which could be extremely complicated and difficult to fully understand. Nevertheless, the effect of the detector's decoherence in time can be formally modeled by reducing the switch-on period of the switching function $\chi(t)$.
Prescribing the tractable form \eqref{chi} for $\chi(t)$ and studying the limit $v_0\Delta_\chi\ll\Delta$, we have shown in \eqref{reduced to classical} that the collective result of an ensemble of Unruh-DeWitt-type quantum detectors reproduces the detection probability of an ordinary classical detector, if coherence in time of each individual quantum detector becomes sufficiently short. Equivalently, the accumulated count of individual signals of a single quantum detector also behaves like a low-efficiency classical detector, if its coherence in time is corrupted (for example, by coupling it to a high-precision clock). This affirms the main idea of \emph{decoherence theory} \cite{schlosshauer2005decoherence} that quantum behavior is lost as a result of quantum decoherence.

Our formal model reveals a profound difference between the result measured by ordinary classical detectors and that by Unruh-DeWitt-type quantum detectors, and furthermore demonstrates that the former can be understood as a certain limit of the latter.
This might offers new insight into the measurement problem in quantum mechanics. Particularly, the result of our model seems to support the tenet of \emph{objective-collapse theories} (e.g.\ \cite{ghirardi1985model, ghirardi1986unified, ghirardi1990markov, penrose1996gravity, penrose1998quantum, penrose2014gravitization, jabs2016conjecture}) that a quantum state in superposition is collapsed into a definite state when a certain objective physical threshold is reached (e.g., $v_0\Delta_\chi\ll\Delta$ in our model).

The distinction between classical and quantum detectors also leads to a striking implication. In the setting that $\Delta L\gtrsim6\Delta$, mounting a quantum detector or a classical detector will cause the two-path interference pattern manifested or disappeared, respectively.
Akin to Wheeler's delayed-choice experiment, the choice of mounting a quantum detector or a classical detector can be made \emph{after} the entry of a single-particle pulse into the interferometer.
However, the choice made affects only the detector but apparently makes no change whatsoever upon the two paths.\footnote{\textit{Cf}.\ the result measured by a classical detector with an etalon placed in front of it as discussed earlier for \eqref{P with etalon}.} It is somewhat surprising that the particle's arrival at the detector is affected despite the fact that its passage remain untouched.\footnote{The Aharonov-Bohm (AB) effect \cite{aharonov1959significance} might also be viewed as another example where the particle's arrival at the detector is affected but its passage is untouched. However, even though the magnetic flux is applied outside the passage in the AB effect, the passage is in fact affected as regards the electromagnetic potential.}
In a sense, it is a special kind of manifestation of wave-particle duality that is induced by a change upon the detector, instead of a direct change upon the to-be-measured quantum state.

It should be emphasized that, like Wheeler's delayed-choice experiment, our result of the two-path interference concerns the \emph{single-particle} effect.
The probability of a single-particle pulse registering a signal at $\mathrm{D}_1$ or $\mathrm{D}_2$ as given in \eqref{P result} is measured as the accumulated count of signals by repeating the experiments many times, for each of which a single-particle pulse of the same profile is fired and both the detectors are reset.
To carry out the experiment, the ensemble of emitted single-particle pulses have to be temporally well separated to neglect any contamination from many-particle effects, such as the Hanbury Brown and Twiss (HBT) effect \cite{brown1956test,brown1957interferometry,brown1958interferometry}.

Finally, we remark that our investigation on the two-path interference measured by the Unruh-DeWitt-type quantum detectors is mainly for theoretical and conceptual concerns.
The technology required to conduct the experiment remains extremely challenging, if not impossible, within current reach.
It is not that we do not have a two-state quantum system used as a single-particle detector (e.g., a quantum dot can be served as a single-photon detector \cite{hadfield2009single,eisaman2011single}), but rather the main difficulty is to have long coherence in time that satisfies the condition \eqref{delta t condition}. For a quantum dot coupled to photons \cite{hadfield2009single,eisaman2011single,bimberg1999quantum} as an example, the interaction of coupling sets a upper bound for $\delta t$ and $\delta t$ can only become shorter in the presence of various environmental disturbances.
One has to weaken the coupling strength of photon-detector interaction to make $\delta t$ long enough, but weakening the coupling strength not only yields low detection efficiency but also renders the detection signal less tolerant of environmental noises.
Furthermore, to manifest the two-path interference for the case that $\Delta L$ is considerably larger than $l_c\sim\Delta$, one has to narrow down the spatial width $\Delta$ of the single-particle profile to the extent that
\begin{equation}
6\Delta \lesssim \Delta L \lesssim v_0\delta t
\end{equation}
for a given $\delta t$.
However, for a quantum dot used as a single-photon detector, its application is usually subject to the condition $\Delta\gg v_0\delta t$ as the single-photon pulse typically is to be treated as a monochromatic wave.
It poses an enormous challenge to prolong $\delta t$ of the detector so drastically. Various advanced technologies, especially cryogenic ones \cite{haselden1971cryogenic}, will be required to shield the quantum dot from dissipative interactions with its surroundings to an extreme degree.

%\newpage

\begin{acknowledgments}
This work was initiated as an attempt to rectify a fallacy in the authors' earlier work that was pointed out by an anonymous expert. Substantial improvements were made in response to valuable suggestions from anonymous reviewers of the earlier and current manuscripts. This work was supported in part by the Ministry of Science and Technology, Taiwan under the Grants 107-2112-M-110-003, 107-2119-M-002-031-MY3, 108-2112-M-110-009, 109-2112-M-110-006, 109-2112-M-110-021, 110-2112-M-002-016-MY3, and 110-2112-M-110-015.
\end{acknowledgments}

%\newpage

\appendix

\section{Review and remarks on time-dependent perturbation theory}\label{app:perturbation theory}
One might wonder whether the result \eqref{P result} for the Unruh-DeWitt-type quantum detector can be reproduced by the standard treatment of time-dependent perturbation theory. In the standard procedure to derive the transition rate for $A\rightarrow B$ (e.g., $A$ and $B$ are atomic states) due to an interaction Hamiltonian (e.g., between the atomic states and the radiation field of photons), the states $\ket{A}$ and $\ket{B}$ are treated quantum mechanically. If the $A$-$B$ two-state quantum system is used as a single-particle detector, it apparently will yield a result similar to \eqref{P result} except that the monopole interaction is replaced by a different form (e.g., the magnetic dipole interaction). It turns out, however, the standard treatment in fact assumes the coherence in time of the two-state system to be shorter enough than the time scale of incoming or outgoing pulses while longer enough than $2\pi/\omega$ of the $\omega$ frequency mode, and furthermore it does not take into account interference between different frequency modes. Consequently, the result of the standard treatment is qualitatively different from that of the Unruh-DeWitt-type quantum detector, and we should give up the conventional wisdom of the former in favor of the latter. In this appendix, we scrutinize the subtle yet important difference between the conventional treatment of time-dependent perturbation theory and the unconventional treatment used for the Unruh-DeWitt-type detector in the main text. We follow the lines of Section 2.4 in \cite{sakurai1968advanced} and Section 18.2 in \cite{shankar1994principles} for the standard treatment of time-dependent perturbation theory and refer readers to them for more details.

As a typical example, consider the absorption and emission of photons by a nonrelativistic atomic electron as an example. The interaction Hamiltonian between the electron and the radiation field is given by
\begin{equation}\label{Hint}
H_\mathrm{int} = -\frac{e}{2m}\left(\mathbf{p}\cdot\mathbf{A}(\vx,t) +\mathbf{A}(\vx,t)\cdot\mathbf{p} \right)
+\frac{e^2}{2m}\abs{\mathbf{A}(\vx,t)}^2,
\end{equation}
where $\mathbf{p}$ is the electron's momentum and $\mathbf{A}(\vx,t)$ is the vector potential of the radiation field.
For the absorption and emission of a photon, the quadratic terms $\mathbf{A}\cdot\mathbf{A}$ does not contribute in the lowest order, since it changes the total number of photons by $0$ or $\pm2$. The matrix element corresponding to the transition from an atomic state $\ket{A}$ to another state $\ket{B}$ by absorbing a photon characterized by momentum $\vk$ and polarization $\alpha$ is given by
\begin{eqnarray}\label{Hint abs}
&&\bra{B,n_{\vk,\alpha}\!-\!1} H_\mathrm{int} \ket{A,n_{\vk,\alpha}} = -\frac{e}{m} \frac{1}{\sqrt{2\omega V}}\nonumber\\
&& \qquad \mbox{}\times \bra{B,n_{\vk,\alpha}\!-\!1}
a_{\vk,\alpha} e^{i(\kx-\omega t)} \mathbf{p}\cdot\boldsymbol{\epsilon}^{(\alpha)}
\ket{A,n_{\vk,\alpha}} \nonumber\\
&=& -\frac{e}{m} \sqrt{\frac{n_{\vk,\alpha}}{2\omega V}}\,
\bra{B}e^{i\kx}\mathbf{p}\cdot\boldsymbol{\epsilon}^{(\alpha)}\ket{A}e^{-i\omega t},
\end{eqnarray}
and similarly that by emitting a photon is given by
\begin{eqnarray}\label{Hint emis}
&&\bra{B,n_{\vk,\alpha}\!+\!1} H_\mathrm{int} \ket{A,n_{\vk,\alpha}} = -\frac{e}{m} \frac{1}{\sqrt{2\omega V}} \nonumber\\
&& \qquad \mbox{}\times \bra{B,n_{\vk,\alpha}\!+\!1}
a_{\vk,\alpha}^\dag e^{-i(\kx-\omega t)} \mathbf{p}\cdot\boldsymbol{\epsilon}^{(\alpha)}
\ket{A,n_{\vk,\alpha}} \nonumber\\
&=& -\frac{e}{m} \sqrt{\frac{n_{\vk,\alpha}\!+\!1}{2\omega V}}\,
\bra{B}e^{-i\kx}\mathbf{p}\cdot\boldsymbol{\epsilon}^{(\alpha)}\ket{A}e^{i\omega t},
\end{eqnarray}
where $\omega$ is shorthand for $\omega_\vk$.
If we apply a monochromatic radiation field given by
\begin{equation}\label{monochromatic radiation}
\mathbf{A}(\vx,t) =\mathbf{A}_0\, e^{i(\kx-\omega t)},
\end{equation}
the amplitude $\mathbf{A}_0$ can be viewed as given by a fixed number $n_{\vk,\alpha}$ as
\begin{subequations}
\begin{eqnarray}
\mathbf{A}_0^{(\mathrm{abs})} &=& \sqrt{\frac{n_{\vk,\alpha}}{2\omega V}} \,\boldsymbol{\epsilon}^{(\alpha)}, \\
\mathbf{A}_0^{(\mathrm{emis})} &=& \sqrt{\frac{n_{\vk,\alpha}\!+\!1}{2\omega V}} \,\boldsymbol{\epsilon}^{(\alpha)},
\end{eqnarray}
for absorption and emission (including spontaneous emission with $n_{\vk,\alpha}=0$), respectively.
\end{subequations}

For a multiple-state quantum system, the wave function of the system can be expanded as
\begin{equation}
\psi(\vx,t) = \sum_{n} c_n(t) u_n(\vx)e^{-iE_nt},
\end{equation}
where $\ket{u_n}$ is the energy eigenstate of the unperturbed Hamiltonian $H_0:=\sum_{n=A,B}E_n\ket{u_n}\bra{u_n}$ with energy $E_n$. If the system is subject to a time-dependent perturbation characterized by the Hamiltonian $H_I(t)$, the transition between different states can be induced.
Accordingly to the time-dependent perturbation theory, we have
\begin{equation}
\dot{c}_m = \sum_n -i\bra{m}H_I(t)\ket{n} e^{i(E_m-E_n)t} c_n(t).
\end{equation}
Particularly, for an $A$-$B$ two-state system, if the initial state is given by $\psi(\vx,t=0)=u_A(\vx)$, i.e., $u_A(t=0)=1$ and $u_B(t=0)=0$, provided that $H_I(t)$ is weak enough, we can approximate $u_B(t)$ up to the first order as
\begin{equation}\label{uB(t)}
u_B(t) = -i\int_0^t dt' \bra{B} H_I(t') \ket{A}\, e^{i\Delta E t'},
\end{equation}
where $\Delta E=E_B-E_A$.
In the case of monochromatic radiation, $\bra{B} H_I(t') \ket{A}$ is given by \eqref{Hint abs} and \eqref{Hint emis}, and hence
\begin{equation}
\bra{B}H_I(t)\ket{A} = \bra{B}H'\ket{A} e^{\mp\omega t},
\end{equation}
where $H'$ is a time-independent operator and ``$\pm$'' is for absorption and emission, respectively.
Consequently, we have
\begin{equation}\label{uB(t) 2}
u_B(t) = -i\bra{B} H' \ket{A} \int_0^t dt' e^{i(\Delta E\mp\omega) t'},
\end{equation}
which follows
\begin{equation}
\abs{u_B(t)}^2 = \abs{\bra{B} H' \ket{A}}^2 \left(\frac{\sin\left[(\Delta E\mp\omega)t/2\right]}{(\Delta E\mp\omega)t/2}\right)^2 t^2.
\end{equation}
Since the function $\sin^2x/x^2$ is peaked at $x=0$ and has a width $\delta x\simeq\pi$, the quantity $\abs{u_B(t)}^2$ is appreciable if
\begin{equation}
\abs{(\Delta E\mp\omega)t/2} \lesssim \pi,
\end{equation}
or equivalently
\begin{equation}\label{Delta E and omega}
\Delta E = \pm\omega \left(1\pm\frac{2\eta}{\omega t}\right), \quad\text{with}\ \abs{\eta}\lesssim\pi.
\end{equation}
If $t$ is small, the energy difference $\Delta E$ shows no particular preference for $\omega$. On the other hand, when $t\gtrsim 2\pi/\omega$, $\Delta E$ starts to favor $\omega$ that is close to the energy gap, i.e., $\omega \approx \pm \Delta E$ (see p.~482 in \cite{shankar1994principles} for more discussions).
If $t\gg2\pi/\omega$, we can take the formal limit $t\rightarrow\infty$ and use the identity
\begin{equation}\label{sinc to delta}
\lim_{t\rightarrow\infty}\frac{t}{\pi}\frac{\sin\alpha t}{\alpha t}=\delta(\alpha).
\end{equation}
Consequently, we have
\begin{equation}\label{uB(t) square}
\abs{u_B(t)}^2 = (2\pi)^2\abs{\bra{B} H' \ket{A}}^2 \left(\delta(\Delta E\mp\omega)\right)^2.
\end{equation}
Note that in the first-order perturbation theory, \eqref{uB(t)} is legitimate for longer time $t$ if the perturbation Hamiltonian $H'$ is weaker. However, no matter how weak $H'$ is, rigorously speaking, $t$ cannot be taken to infinity. The $t\rightarrow\infty$ limit is only a \emph{formal} prescription, which gives rise to the divergent behavior of \eqref{uB(t) square} in direct contradiction to the condition $\abs{u_B(t)}^2\leq1$ and thus has to be regularized to make physical sense. To regularize the divergent behavior, we consider
\begin{eqnarray}
\left(\delta(\Delta E\mp\omega)\right)^2 &=& \lim_{T\rightarrow\infty}\delta(\Delta E\mp\omega)\frac{1}{2\pi}\int_{-T/2}^{T/2} e^{i(\Delta E\mp\omega)t}dt \nonumber\\
&\sim& \delta(\Delta E\mp\omega)\frac{T}{2\pi}.
\end{eqnarray}
Therefore, as far as the transition amplitude $\abs{u_B(t)}^2$ averaged over a long enough period $T$ is concerned, we can still sensibly talk about the \emph{transition rate} given by
\begin{equation}\label{R A to B}
\mathcal{R}_{A\rightarrow B}=\frac{\abs{u_B(t)}^2}{T} = 2\pi\abs{\bra{B} H' \ket{A}}^2 \delta(\Delta E\mp\omega),
\end{equation}
which is independent of $t$ and $T$.
The delta function $\delta(\Delta E\mp\omega)$ appearing in the transition rate is the celebrated \emph{Fermi's golden rule}.

What happens if the radiation field is not a monochromatic plane wave? That is, the radiation field is not given by \eqref{monochromatic radiation} but by a wave-package wave in a generic form
\begin{equation}
\mathbf{A}(\vx,t) = \sum_\alpha\int \frac{d^3k}{\sqrt{V}} \frac{\tilde{f}^{(\alpha)}(\vk)}{\sqrt{2\omega_\vk}}\, \boldsymbol{\epsilon}^{(\alpha)}_\vk\, e^{i(\kx-\omega_\vk t)},
\end{equation}
where $\tilde{f}^{(\alpha)}(\vk)$ is the amplitude for the $(\vk,\alpha)$ mode.
The standard treatment is to sum up transition rates for different modes; i.e., the total transition rate is given by
\begin{eqnarray}\label{R total}
&&\frac{\abs{u_B(t)}^2}{T} \approx \sum_\alpha\int \frac{d^3k}{V} \abs{\tilde{f}^{(\alpha)}(\vk)}^2 \mathcal{R}_{A\rightarrow B}(\vk,\alpha) \\
&=& 2\pi \sum_\alpha\int \frac{d^3k}{V} \abs{\tilde{f}^{(\alpha)}(\vk)}^2 \abs{\bra{B} H'(\vk,\alpha) \ket{A}}^2 \delta(\Delta E\mp\omega_\vk),\nonumber
\end{eqnarray}
where $\mathcal{R}_{A\rightarrow B}(\vk,\alpha)$ for a specific $\vk$ and $\alpha$ is given by \eqref{R A to B}, and $T$ is the time span of the wave-package wave. It should be remarked that in \eqref{R total} contributions from different $(\vk,\alpha)$ modes are summed up \emph{additively} rather than \emph{interferentially}. In other words, we neglect interference between any two modes.

The standard treatment yields correct results for most experimental settings. For example, in the experiment of absorption spectroscopy, an incident light of multiple wavelengths is applied to an analyte, and the absorption spectrum is obtained by comparing the attenuation of the light transmitted through the analyte with the original incident light.
Because an analyte is composed of a huge number (typically many moles) of molecules that are decoherent with one another, the interaction between a pair of a photon and a molecule is independent of another photon-molecule pair. Therefore, as far as the spectral lines of the analyte \emph{as a whole} is concerned, we shall sum up the transition rates additively without taking into account any interference between different $(\vk,\alpha)$ modes, even if the incident light is given by a coherent light source.
Although we consider the quantum states $\ket{A}$ and $\ket{B}$ of a \emph{single} quantum system to derive the transition amplitude \eqref{uB(t) square}, the resulting transition rate given by \eqref{R A to B} or \eqref{R total} is usually used for the experimental setting of a huge \emph{ensemble} of such quantum systems.

What if the experimental setting is truly of a \emph{single} quantum system? For example, consider a single quantum dot used as a single-photon detector (e.g.\ see \cite{eisaman2011single}). In this case, if we conduct the experiment repeatedly and measure the accumulated counts of signals, do we still use \eqref{R total} or we have to take into account of interference between different $(\vk,\alpha)$ modes? The answer depends on how long the quantum system can remain coherent in time.
If the incident light is given by a coherent wave packet, the amplitude of the light takes the form similar to \eqref{wave packet}, and typically we have $\lambda_0\equiv 2\pi/\abs{\vk_0}\ll\Delta_i$ (i.e., the wavelength of the incident light is assumed to be much shorter than the spatial width of the wave packet). If the coherent time $\delta t$ of the single-photon detector satisfies the condition $\lambda_0/\abs{\vv_0}\approx2\pi/\omega_0\ll\delta t\ll\Delta_i/\abs{\vv_0}$, then we can still prescribe the formal limit $t\rightarrow\infty$ in \eqref{uB(t)} since $t\gtrsim 2\pi/\omega$, and use \eqref{R total} with $\tilde{f}^{(\alpha)}(\vk)$ given in the form of \eqref{f tilde} for the transition rate without considering any interference between different $(\vk,\alpha)$ modes.

On the other hand, it could be extremely difficult as discussed in \secref{sec:discussion} to make the coherent time of the detector long enough such that $\lambda_0/\abs{\vv_0}\ll\Delta_i/\abs{\vv_0}\lesssim\delta t$. If we manage to achieve it, we have to take into account the interference between different $(\vk,\alpha)$ modes.
Consider a single-photon pulse state $\ket{\Psi_0}$ given by
\begin{equation}\label{single-photon pulse Psi0}
\ket{\Psi_0} = \sum_\alpha\int\frac{d^3k}{\sqrt{V}}\frac{\tilde{f}^{(\alpha)}(\vk)}{\sqrt{2\omega_\vk}}\, a_{\vk,\alpha}^\dag \boldsymbol{\epsilon}^{(\alpha)*}_\vk \ket{0},
\end{equation}
which is in a form similar to \eqref{Psi0 in k}. If we apply $\ket{\Psi_0}$ to the detector, the interaction Hamiltonian \eqref{Hint} will induce the detector to undergo the transition from $\ket{A}$ to $\ket{B}$ by absorbing a photon from $\ket{\Psi_0}$. The correspondingly matrix element of the transition is given by
\begin{eqnarray}\label{Hint abs new}
&&\bra{B,0} H_\mathrm{int}(t) \ket{A,\Psi_0} \equiv \bra{B,0} H_\mathrm{int}(t=0) \ket{A,\Psi_0(t)} \nonumber\\
&=&\sum_{\alpha,\alpha'}\int d^3kd^3k'\, \tilde{f}^{(\alpha')}(\vk)
\frac{1}{\sqrt{2\omega_\vk V}}\frac{1}{\sqrt{2\omega_{\vk'} V}}\frac{-e}{m}\nonumber\\
&& \mbox{}\times \bra{0,B}
a_{\vk,\alpha} e^{i(\kx-\omega_\vk t)} \mathbf{p}\cdot\boldsymbol{\epsilon}^{(\alpha)}_\vk
a_{\vk,\alpha'}^\dag\boldsymbol{\epsilon}^{(\alpha')*}_{\vk'} \ket{0,A} \nonumber\\
&=& \sum_\alpha\int d^3k \, \tilde{f}^{(\alpha)}(\vk) \nonumber\\
&& \quad\mbox{}\times\frac{-e}{m} \frac{1}{\sqrt{2\omega_\vk V}}\,
\bra{B}e^{i\kx}\mathbf{p}\cdot\boldsymbol{\epsilon}^{(\alpha)}_\vk\ket{A}\,e^{-i\omega_\vk t},\quad
\end{eqnarray}
where we have used $[a_{\vk,\alpha},a_{\vk,\alpha'}^\dag] =\delta(\vk-\vk')\delta_{\alpha,\alpha'}$, $\boldsymbol{\epsilon}^{(\pm)}\!\cdot\boldsymbol{\epsilon}^{(\pm)*} =-\boldsymbol{\epsilon}^{(\pm)}\!\cdot\boldsymbol{\epsilon}^{(\mp)}=1$, and $\boldsymbol{\epsilon}^{(\pm)}\!\cdot\boldsymbol{\epsilon}^{(\mp)*} =-\boldsymbol{\epsilon}^{(\pm)}\!\cdot\boldsymbol{\epsilon}^{(\pm)}=0$.
The matrix element \eqref{Hint abs new} is the sum of \eqref{Hint abs} with $n_{\vk,\alpha}=1$ over different $\vk$ modes interfered with one another. Instead of \eqref{uB(t) 2}, the first-order perturbation theory now yields
\begin{eqnarray}\label{uB(t) unconventional}
u_B(t) &=& -i \sum_\alpha\int d^3k \, \tilde{f}^{(\alpha)}(\vk) \bra{B} H'(\vk,\alpha) \ket{A}\nonumber\\
&& \quad\qquad \mbox{}\times  \int_0^t dt' e^{i(\Delta E-\omega_\vk) t'},
\end{eqnarray}
which, in the limit $t\rightarrow\infty$, leads to
\begin{eqnarray}\label{uB(t) square unconventional}
&&\abs{u_B(t)}^2 \\
&=& \bigg|2\pi\sum_\alpha\int d^3k \, \tilde{f}^{(\alpha)}(\vk) \bra{B} H'(\vk,\alpha) \ket{A} \delta(\Delta E-\omega_\vk)\bigg|^2 \nonumber
\end{eqnarray}
by the identity \eqref{sinc to delta} again.
If the $\ket{\Psi_0}$ is given as a monochromatic plane-wave state with $\tilde{f}(\vk)=\delta(\vk-\vk_0)$, \eqref{uB(t) square unconventional} is reduced back to \eqref{uB(t) square}, which is divergent and needs to be regularized. On the other hand, if $\ket{\Psi_0}$ is given as a wave-package state with $\tilde{f}(\vk)$ given as \eqref{f tilde}, \eqref{uB(t) square unconventional} is finite and accurately represents the \emph{transition probability} provided that $H'$ is weak enough. In the latter case, we can directly calculate the transition probability without the regularization appealing to the transition rate.
It should be emphasized that \eqref{uB(t) square unconventional} is different from \eqref{R total} not only in the sense that they are of different dimensions (probability vs.\ probability per unit time) but more importantly in the fact that \eqref{uB(t) square unconventional} takes into account the interference between different $(\vk,\alpha)$ modes that satisfy $\Delta E=\omega_\vk$ whereas \eqref{R total} does not.

Now, instead of \eqref{single-photon pulse Psi0}, consider the case that $\ket{\Psi_0}$ is given as a single-photon state that is made of two coherent wave packages separated by $\Delta\mathbf{L}$ with $\abs{\Delta\mathbf{L}}\gtrsim6\Delta_i$ as depicted in \figref{fig:functions}. The amplitude $\tilde{f}^{(\alpha)}(\vk)$ of $\ket{\Psi_0}$ takes the form
\begin{equation}\label{f tilde for two pulses}
\tilde{f}^{(\alpha)}(\vk) = \frac{1}{\sqrt{2}}\tilde{f}^{(\alpha)}_0(\vk) + \frac{e^{i\theta}}{\sqrt{2}}\,e^{i\vk\cdot\Delta\mathbf{L}} \tilde{f}^{(\alpha)}_0(\vk),
\end{equation}
where $\tilde{f}^{(\alpha)}_0(\vk)$ is the $(\vk,\alpha)$ mode amplitude of a single wave package, and $e^{i\theta}$ reprents an extra phase difference between the two wave packages.
This time, do we have to take into account the interference \emph{between} the two wave package as well as that \emph{within} each wave package? The answer depends again on how long the quantum detector remains coherent in
time.
If $\Delta_i/\abs{\vv_0}\lesssim\delta t\lesssim\abs{\Delta\mathbf{L}}$, the two wave packages are to be viewed as two independent pulses, and the resulting transition probability is just the sum of each contribution, i.e.,
\begin{eqnarray}
&&\abs{u_B(t)}^2 = \abs{u_B(t)}^2_\mathrm{1st\ package} + \abs{u_B(t)}^2_\mathrm{2nd\ package}\qquad\quad \\
&=& \bigg|2\pi\sum_\alpha\int d^3k \, \tilde{f}^{(\alpha)}_0(\vk) \bra{B} H'(\vk,\alpha) \ket{A} \delta(\Delta E-\omega_\vk)\bigg|^2, \nonumber
\end{eqnarray}
which is identical to that of a single wave-package and does not manifest any interference between the two wave packages.
On the other hand, if $\abs{\Delta\mathbf{L}}\lesssim\delta t$, we shall directly substitute \eqref{f tilde for two pulses} into \eqref{uB(t) square unconventional}, and the resulting transition probability is
\begin{eqnarray}\label{uB(t) square for two pulses}
\abs{u_B(t)}^2
&=& 2\pi^2\bigg|\sum_\alpha\int d^3k\, (1+e^{i(\theta+\vk\cdot\Delta\mathbf{L})})\tilde{f}^{(\alpha)}_0(\vk) \nonumber\\
&& \qquad\;\mbox{}\times\bra{B} H'(\vk,\alpha) \ket{A} \delta(\Delta E-\omega_\vk)\bigg|^2,\qquad
\end{eqnarray}
which does manifest the interference between the two coherent wave packages.

In summary, depending on the exact experimental setting, we may or may not have to modify the standard treatment by taking into account the interference between different $(\vk,\alpha)$ modes. In the main text, we assume the single-particle detector to have coherence in time longer than the separation between the two wave packages, and it is crucial to consider the inter-package interference. Note that \eqref{A'} is in a form similar to \eqref{uB(t) unconventional} except that the filed $\mathbf{A}(\mathbf{x},t)$ is replaced by the scalar field $\phi(x,t)$ and the magnetic dipole (M1) interaction is replaced by the Unruh-DeWitt monopole interaction. The simplicity of the Unruh-DeWitt model enables us to explicitly calculate the transition probability. The inter-package interference factor $1+e^{i(\theta+\vk\cdot\Delta\mathbf{L})}$ in \eqref{uB(t) square for two pulses} is responsible for the modulation in response to $\theta$ in \eqref{P result}.
In \secref{sec:quantum vs classical}, using the Unruh-DeWitt model, we investigate in depth how the interference between the two wave packages gradually loses its significance when the coherence in time of the single-particle detector becomes shorter and shorter.

\section{Detailed derivations of various equations}
%\newpage
Detailed derivations of various equations are provided here.

\subsection{Detailed derivation of \eqref{Psi t 2}}\label{app:Psi t 2}
Substituting \eqref{Taylor expansion} into \eqref{Psi t a}, we have
\begin{eqnarray}
&&\ket{\Psi(t)} \nonumber\\
&\approx& e^{-i(\omega_0-\vv_0\cdot\vk_0)t}\int d^nx f(\vx) \int\frac{d^nk}{(2\pi)^n} \frac{e^{-i\vk\cdot(\vx+\vv_0t)}}{\sqrt{2\omega_\vk}} a_\vk^\dag\ket{0}\nonumber\\
&=&  e^{-i\delta\omega_0t}\int d^nx f(\vx-\vv_0t) \int\frac{d^nk}{(2\pi)^n} \frac{e^{-i\kx}}{\sqrt{2\omega_\vk}} a_\vk^\dag\ket{0}, \nonumber\\
&\equiv&  e^{-i\delta\omega_0t}\int d^nx f(\vx-\vv_0t) \phi(\vx)\ket{0}.
\end{eqnarray}

\subsection{Detailed derivation of \eqref{AL result}}\label{app:AL result}
Performing the change of variables $y'=L+y-v_0t$ and $t'=t$ upon \eqref{AL} and noting that the Jacobian determinant $\abs{\partial(y',t')/\partial(y,t)}=1$, we then have
\begin{eqnarray}
\mathcal{A}_L &=&
\kappa \bra{E} \mu(0) \ket{E_0}
\int_{-\infty}^\infty dt'\int_{-\infty}^\infty dy' \int_{-\infty}^\infty\frac{dk}{2\pi}\frac{1}{2\omega_k} \nonumber\\
&&\quad\mbox{}\times e^{i(\Delta E-\delta\omega_0-kv_0)t'} e^{-ik(y'-L)} f(y') \nonumber\\
&=&
\kappa \bra{E} \mu(0) \ket{E_0}
\int_{-\infty}^\infty dy'
\int_{-\infty}^\infty\frac{dk}{2\omega_k} \nonumber\\
&&\quad\mbox{}\times \delta(\Delta E-\delta\omega_0-kv_0) e^{-ik(y'-L)} f(y') \nonumber\\
&=&\frac{\kappa \bra{E} \mu(0) \ket{E_0}}{2\omega_{k_*}\abs{g'(k_*)}}
e^{ik_*L}\int_{-\infty}^\infty dy f(y)e^{-ik_*y} \nonumber\\
&=:& e^{ik_*L} \mathcal{A}_0(k_*),
\end{eqnarray}
where the function $g(k)$ is defined as
\begin{eqnarray}\label{g(k)}
g(k)&:=&\Delta E-\delta\omega_0-kv_0 \\
&=& \Delta E - \omega_0 + v_0k_0 - kv_0 \equiv v_0(k_*-k_0), \nonumber
\end{eqnarray}
whose root is denoted as
\begin{equation}%\label{k*}
k_* = \frac{1}{v_0}\left(\Delta E - \omega_0 + v_0k_0\right),
\end{equation}
and $\mathcal{A}_0(k_*)$ is given by \eqref{A0}.

\subsection{Detailed derivation of \eqref{AL result chi}}\label{app:AL result chi}
Substituting \eqref{chi} into \eqref{AL} and performing the change of variables $z=L+y-v_0t$ and $t'=t$, we then have
\begin{eqnarray}%\label{AL result chi}
\mathcal{A}_L &=&
\kappa \bra{E} \mu(0) \ket{E_0}
\int_{-\infty}^\infty dt'
\int_{-\infty}^\infty dz
\int_{-\infty}^\infty \frac{dk}{2\pi}\frac{1}{2\omega_k} \nonumber\\
&&\quad\mbox{}\times e^{i(\Delta E-\delta\omega_0-kv_0)t'}
                     e^{-(t-t_\chi)^2/2\Delta_\chi^2} \nonumber\\
&&\quad\mbox{}\times e^{-ik(z-L)} f(z).
\end{eqnarray}
Applying the Gaussian integral formula
\begin{equation}\label{Gaussian formula}
\int_{-\infty}^\infty dx\, e^{-ax^2+bx+c}=\sqrt{\frac{\pi}{a}}\, e^{\frac{b^2}{4a}+c}
\end{equation}
to the integration over $t'$:
\begin{equation}
\mathcal{I}(k)
= \int_{-\infty}^\infty dt' e^{i(\Delta E-\delta\omega_0-kv_0)t'} e^{-(t'-t_\chi)^2/2\Delta_\chi^2},
\end{equation}
we obtain
\begin{equation}
 \mathcal{I}(k)=\sqrt{2\pi}\,\Delta_\chi\, e^{ig(k)t_\chi-8\Delta_\chi^2g(k)^2},
\end{equation}
where $g(k)$ is defined in \eqref{g(k)}. Consequently, we have
\begin{eqnarray}%\label{AL result chi}
\mathcal{A}_L &=&
\sqrt{2\pi}\,\Delta_\chi\kappa \bra{E} \mu(0) \ket{E_0}
\int_{-\infty}^\infty dz \int_{-\infty}^\infty\frac{dk}{2\pi}\frac{1}{2\omega_k} \nonumber\\
&&\quad\mbox{}\times e^{ig(k)t_\chi-8\Delta_\chi^2g(k)^2} e^{-ik(z-L)} f(z).
\end{eqnarray}

\subsection{Detailed derivation of \eqref{deriving average AAstar}}\label{app:deriving average AAstar}
The ensemble average of $\mathcal{A}_{L_1} \mathcal{A}^*_{L_2}$ is proportional to
\begin{eqnarray}
&&\frac{1}{\Delta_\chi}\int_{-\infty}^\infty dt_\chi \mathcal{A}_{L_1} \mathcal{A}^*_{L_2}\nonumber\\
&=& 2\pi \Delta_\chi \abs{\kappa \bra{E} \mu(0) \ket{E_0}}^2
\int_{-\infty}^\infty\frac{dk}{2\pi}\frac{1}{2\omega_k}
\int_{-\infty}^\infty\frac{dk'}{2\pi}\frac{1}{2\omega_{k'}}
\nonumber\\
&&\quad\mbox{}\times \int_{-\infty}^\infty dt_\chi e^{i(g(k)-g(k'))t_\chi} e^{-8\Delta_\chi^2(g(k)^2+g(k')^2)} \nonumber\\
&&\quad\mbox{}\times \int_{-\infty}^\infty dz\, e^{-ik(z-L_1)} f(z) \int_{-\infty}^\infty dz' e^{ik'(z'-L_2)} f(z')^* \nonumber\\
%-------
&=&\left(2\pi\right)^2 \Delta_\chi\abs{\kappa \bra{E} \mu(0) \ket{E_0}}^2
\int_{-\infty}^\infty\frac{dk}{2\pi}\frac{1}{2\omega_k}
\int_{-\infty}^\infty\frac{dk'}{2\pi}\frac{1}{2\omega_{k'}}
\nonumber\\
&&\quad\mbox{}\times \delta(g(k)-g(k'))\, e^{-8\Delta_\chi^2(g(k)^2+g(k')^2)} \nonumber\\
&&\quad\mbox{}\times \int_{-\infty}^\infty dz\, e^{-ik(z-L_1)} f(z) \int_{-\infty}^\infty dz' e^{ik'(z'-L_2)} f(z')^* \nonumber\\
%-------
&=&\left(2\pi\right)^2 \Delta_\chi\abs{\kappa \bra{E} \mu(0) \ket{E_0}}^2
   \int_{-\infty}^\infty\frac{dk}{(2\pi)^2}\frac{1}{(2\omega_k)^2}\frac{1}{\abs{g'(k)}} \nonumber\\
&& \quad\mbox{}\times e^{-16\Delta_\chi^2g(k)^2}e^{-ik\Delta L} \nonumber\\
&& \quad \mbox{}\times
   \int_{-\infty}^\infty dz\, e^{-ikz} f(z)
   \int_{-\infty}^\infty dz' e^{ikz'} f(z')^* \nonumber\\
&=&2\pi \Delta_\chi\abs{\kappa \bra{E} \mu(0) \ket{E_0}}^2
\int_{-\infty}^\infty dk\frac{1}{(2\omega_k)^2}\frac{1}{\abs{g'(k)}}\nonumber\\
&&\quad\mbox{}\times e^{-16\Delta_\chi^2g(k)^2} e^{-ik\Delta L} \tilde{f}(k) \tilde{f}(k)^* \nonumber\\
&=:&\sqrt{\frac{\pi}{2}}\,\Delta\Delta_\chi\abs{\kappa \bra{E} \mu(0) \ket{E_0}}^2
\int_{-\infty}^\infty dk\frac{e^{-h(k)-ik\Delta L}}{\omega_k^2\abs{g'(k)}},\nonumber\\
\end{eqnarray}
where $h(k)$ is defined in \eqref{def h}.

\newpage

%\bibliography{ref}
%\bibliographystyle{ieeetr}

\end{document}